\documentclass[fleqn,usenatbib]{mnras}

\usepackage{newtxtext,newtxmath}
\usepackage[T1]{fontenc}
\usepackage{ae,aecompl}

\DeclareRobustCommand{\VAN}[3]{#2}
\let\VANthebibliography\thebibliography
\def\thebibliography{\DeclareRobustCommand{\VAN}[3]{##3}\VANthebibliography}
\usepackage{soul}
\usepackage{xcolor}
\usepackage{todonotes}
\usepackage{commath}

\newcommand{\optimize}{\texttt{Scipy.optimize.fmin\_cg()}}
\newcommand{\sparse}{\texttt{sparse\_2level()}}

\newcommand{\scipy}{\texttt{Scipy}}
\newcommand{\corrcal}{\texttt{CorrCal}}
\newcommand{\omnical}{\texttt{OmniCal}}
 
\newcommand{\iu}{\mathrm{i}\mkern1mu}

\def\I{\textbf{\textsf{I}}}
\def\u{\textbf{\textit{u}}}
\def\d{\textbf{\textit{d}}}
\def\n{\textbf{\textit{n}}}
\def\G{\textbf{\textsf{G}}}
\def\v{\textbf{\textit{v}}}
\def\m{\textbf{\textit{m}}}
\def\N{\textbf{\textsf{N}}}
\def\C{\textbf{\textsf{C}}}
\def\S{\textbf{\textsf{S}}}
\def\R{\textbf{\textsf{R}}}
\def\A{\textbf{\textsf{A}}}
\def\b{\textbf{\textit{b}}}
\def\D{\textbf{\textsf{D}}}
\def\p{\textbf{\textsf{p}}}
\def\s{\textbf{\textsf{s}}}
\def\r{\textbf{\textsf{r}}}

\def\v{\textbf{\textsf{v}}}

\def\B{\textbf{\textsf{B}}}
\def\ur{\hat{\textbf{\textit{r}}}}
\def\hatx{\hat{\textbf{\textit{x}}}}
\def\t1{\boldsymbol{\theta}}
\def\Sig1{\boldsymbol{\Sigma}}

\newcommand{\reffg}[1]{Fig.~\ref{#1}}

\newcommand{\refeq}[1]{Eq.~(\ref{#1})}

\RequirePackage[normalem]{ulem} 
\RequirePackage{color}\definecolor{RED}{rgb}{1,0,0}\definecolor{BLUE}{rgb}{0,0,1} 
\RequirePackage{listings} 
\RequirePackage{color} 
\lstdefinelanguage{DIFcode}{ 
	moredelim=[il][\color{red}\sout]{\%DIF\ <\ }, 
	moredelim=[il][\color{blue}\uwave]{\%DIF\ >\ } 
} 
\lstdefinestyle{DIFverbatimstyle}{ 
	language=DIFcode, 
	basicstyle=\ttfamily, 
	columns=fullflexible, 
	keepspaces=true 
} 
\lstnewenvironment{DIFverbatim}{\lstset{style=DIFverbatimstyle}}{} 
\lstnewenvironment{DIFverbatim*}{\lstset{style=DIFverbatimstyle,showspaces=true}}{} 

\usepackage{graphicx}	
\usepackage{amsmath}	
\usepackage{amssymb}	

\title[The Correlation Calibration]{The Correlation Calibration of PAPER-64 data}
\vspace{-30pt}
\author[Tamirat et al.]{Tamirat G. Gogo,$^{1,2}$
Yin-Zhe Ma$^{1,2}$\thanks{Corresponding author: ma@ukzn.ac.za},
Piyanat Kittiwisit$^{1,2}$,
Jonathan L. Sievers$^{3, 1}$,
Aaron R. Parsons$^{4, 5}$,
\newauthor
Jonathan C. Pober$^{6}$,
Daniel C. Jacobs$^{7}$,
Carina Cheng $^{4}$,
Matthew Kolopanis$^{7}$,
Adrian Liu$^{8, 9}$,
\newauthor
Saul A. Kohn$^{10}$,
James E. Aguirre$^{10}$,
Zaki S. Ali,$^{5}$,
Gianni Bernardi$^{11, 12, 13}$,
Richard F. Bradley$^{14, 15,16}$,
\newauthor
David R. DeBoer$^{5}$,
Matthew R. Dexter$^{5}$,
Joshua S. Dillon$^{4}$, 
Pat Klima$^{15}$, 
David H. E. MacMahon$^{5}$, \newauthor
David F. Moore$^{10}$, 
Chuneeta D. Nunhokee$^{4}$,
William P. Walbrugh$^{13}$,
and Andre Walker$^{11}$\\
$^{1}$ School of Chemistry and Physics, University of KwaZulu-Natal, Westville Campus, Private Bag X54001, Durban, 4000, South Africa\\
$^{2}$ NAOC-UKZN Computational Astrophysics Centre (NUCAC), University of KwaZulu-Natal, Durban, 4000, South Africa\\
$^{3}$ Department of Physics, McGill University, 3600 University Street Montreal, QC H3A 2T8, Canada\\
$^{4}$ Department of Astronomy, UC Berkeley, Berkeley, CA 94720, USA\\
$^{5}$ Radio Astronomy Laboratory, UC Berkeley, Berkeley, CA 94720, USA\\
$^{6}$ Department of Physics, Brown University, Providence, RI 02912, USA\\
$^{7}$ School of Earth and Space Exploration, Arizona State University, Tempe, AZ 85287, USA\\
$^{8}$ Department of Physics and McGill Space Institute, McGill University, Montreal, QC, Canada\\
$^{9}$ CIFAR Azrieli Global Scholar, Gravity \& the Extreme Universe Program, Canadian Institute for Advanced Research, 661 University\\
Ave., Suite 505, Toronto, Ontario M5G 1M1, Canada\\
$^{10}$ Department of Physics and Astronomy, U. Penn., Philadelphia PA, USA\\
$^{11}$ INAF-Istituto di Radioastronomia, via Gobetti 101, 40129, Bologna, Italy\\
$^{12}$ Department of Physics and Electronics, Rhodes University, PO Box 94, Grahamstown, 6140, South Africa\\
$^{13}$ South African Radio Astronomy Observatory, Black River Park, 2 Fir Street, Observatory, Cape Town, 7925, South Africa\\
$^{14}$ Department of Electrical and Computer Engineering, U. Virginia, Charlottesville VA, USA\\
$^{15}$ National Radio Astronomy Obs., Charlottesville VA, USA\\
$^{16}$ Department of Astronomy, U. Virginia, Charlottesville VA, USA\\
}

\date{Accepted XXX. Received YYY; in original form ZZZ}

\pubyear{2021}
\begin{document}
\label{firstpage}
\pagerange{\pageref{firstpage}--\pageref{lastpage}}
\maketitle

\begin{abstract}
Observation of redshifted 21-cm signal from the Epoch of Reionization (EoR) is challenging due to contamination from the bright foreground sources that exceed the signal by several orders of magnitude. The removal of this very high foreground relies on accurate calibration to keep the intrinsic property of the foreground with frequency. Commonly employed calibration techniques for these experiments are the sky model-based and the redundant baseline-based calibration approaches. However, the sky model-based and redundant baseline-based calibration methods could suffer from sky-modeling error and array redundancy imperfection issues, respectively. In this work, we introduce the hybrid {\it correlation calibration} (\corrcal{}) scheme, which aims to bridge the gap between redundant and sky-based calibration by relaxing redundancy of the array and including sky information into the calibration formalisms. We demonstrate the slight improvement of power spectra, about $-6\%$ deviation at the bin right on the horizon limit of the foreground wedge-like structure, relative to the power spectra before the implementation of \corrcal{} to the data from the Precision Array for Probing the Epoch of Reionization (PAPER) experiment, which was otherwise calibrated using redundant baseline calibration. This small improvement of the foreground power spectra around the wedge limit could be suggestive of reduced spectral structure in the data after \corrcal{} calibration, which lays the foundation for future improvement of the calibration algorithm and implementation method.
\end{abstract}



\section{Introduction}
\label{sec:intro}
The cosmic dark ages have ended at around 100 million years after the big bang, marking the formation of the first structures such as galaxies and stars. These objects started to emit radiation that influences the nearby primordial intergalactic medium (IGM), which predominantly filled with the neutral hydrogen atom. As the number of these very first sources increases,  they emitted sufficient radiation to convert the neutral IGM into the fully ionized medium of the hydrogen atom eventually. The period that marks the transition of neutral IGM into the fully ionized state is known as the Epoch of Reionization (EoR) in the history of the Universe. Thus, investigating the era of EoR will answer the fundamental questions of the formation of the first structures and their properties. 

The measurement of the radiation from the high-redshifted 21-cm emission line is employed to probe the EoR era. This emission line is due to the hyperfine transition of the neutral hydrogen atom in the IGM empowered by the first sources. The information imprinted into the radiation associated with the redshifted 21-cm line can be used as a powerful tool in the modern cosmology to study the ionization state and large-scale temperature fluctuation of IGM, owing to the formation of the first structures \citep[see,  e.g.,][]{Oh2001, Barkan2001, Furlanetto2006, MoralesWyith2010, Pritchard2012}.

A direct probe of the EoR is the tomographic mapping of redshifted 21-cm signals from the neutral hydrogen atom~\citep{Madau1996, Barkan2001, Furlanetto2006, MoralesWyith2010, Pritchard2012}.  Significant progress has made in the past decades. Several experiments dedicated to studying this signal have been built or are being upgraded. The first-generation 21-cm experiments, including the Murchison Widefield Array \citep[MWA;][]{Tingay2013}, the Donald C. Backer Precision Array for Probing the Epoch of Reionization \citep[PAPER;][]{Parsons2010}, the LOw-Frequency ARray \citep[LOFAR;][]{Van_Haarlem_2013}, and the GiantMeter wave Radio Telescope EoR experiment \citep[GMRT;][]{Paciga2013} are already operating and taking data. Observations from these first-generation experiments have set upper limits to the statistical power spectrum of 21-cm brightness temperature from cosmic reionization. In particular, results from MWA, LOFAR, and PAPER observations~\citep[e.g.,][]{Paciga2013, Dillon2014, Patil2017, Cheng2018, Kolopanis2019,Mertens2020} have placed upper limits on the statistical power spectrum of the 21-cm emissions over a broad range of redshifts, providing evidence for the heating of the IGM before reionization. The most recent result from MWA constrains the EoR power spectrum as $\Delta^{2}=(43\,{\rm mK})^{2}=1.8 \times 10^{3}\,{\rm mK}^{2}$ for $k=0.14\,h\,{\rm Mpc}^{-1}$ at $z=6.5$, which is the lowest measurement currently~\citep{Trott20,Rahimi2021}. Lessons learned from these experiments have led to the construction of the second-generation instruments, which include the Hydrogen Epoch of Reionization Array \citep[HERA;][]{DeBoer2017},  the upgraded MWA \citep[MWA-II;][]{Wayth2018}, and the  LOFAR 2.0 Survey~\citep{Edler2021}. With more collecting areas and improved electronics, high-significant measurements of the 21-cm power spectrum are anticipated from these second-generation instruments~\citep{Pober2014,Liu_parson2016, Wayth2018}. 

Detecting the 21-cm signal from EoR is challenging due to the weakness of 21-cm signal comparing to the brightness temperature of the foreground originated from synchrotron radiation from our Galaxy, strong point sources and thermal Bremsstrahlung from the HII region~\citep{Shaver1999, Santos2005, Furlanetto2006, Bernardi2009, Parsons2014}. To disentangle strong foregrounds from weak 21-cm signal, several mitigation techniques have been developed \citep[See e.g.,][]{Wang2006, Liu2009a, Bowman2009a, LiuTegmark2011, Parsons2012a,Dillon2013, Liu2014a, Beardsley2016,Pober2016,Hothi2021}. These techniques, in general, can be categorised into two approaches -- foreground subtraction and foreground avoidance~\citep{Chapman2019}. Foreground subtraction involves modelling of the foreground and subtracting them from the data. In contrast, foreground avoidance completely discards foreground contaminated data from analysis and try to reconstruct the power spectra of EoR in the Fourier regime which are not affected by foreground.

In both foreground mitigation and subtraction, accurate calibration of the array is critical because calibration errors can contaminate the Fourier mode of the EoR power spectrum~\citep{Morales2012,Ewall2017}. Furthermore, simulations in~\citet{Orosz2018} suggested that calibration error introduced by quasi-redundancy in antenna positioning and slight variations of the beam responses between different antennas across the array are significant sources of contamination of the EoR signal.  Even in the limit of perfect antenna positioning and telescope response, the EoR window can still be contaminated if the instrument calibration is susceptible to sky modelling error~\citep{Byrne2019}.

The array configuration of PAPER is designed to enhance sensitivity for a first power-spectrum detection, by measuring same Fourier modes on the sky with a group of redundant baselines. These redundant visibilities need to be calibrated accurately by using the redundancy information of the array. The calibration technique implemented for such kind of array layout is generally known as redundant baseline calibration \citep{Wieringa92,Liu2010}. This technique intends to calibrate measurements from the array of antennas configured on a regular grid spacing with maximum redundancy in baseline distribution and tightly packed with the drift-scan mode of observations. Then one can use statistical tool to calibrate the visibilities for independent baselines.

In this work, we introduce a recently developed calibration scheme called \emph{correlation calibration} (\corrcal{}\footnote{The prototype version of the software located on \url{https://github.com/sievers/corrcal2.git} is being used for this work.}), first proposed in~\citet{S17}. \corrcal{} aims to bridge the gap between sky-based and redundant calibration by incorporating partial sky information and information on the non-redundancies of the array into the calibration solutions. We use this new scheme to re-calibrate data from the 64-element PAPER array (hereafter PAPER64), which is a 9-hour integration data set.~\citet{Ali2015} calibrated this data set with a redundant baseline calibration scheme and published power spectrum results. This work used the same initial calibration techniques employed in \cite{Ali2015}, but have an independent pipeline from that point on.  We will show that \corrcal{} can potentially constrain the  wedge-like structure of foreground in the Fourier space. 

The rest of paper is organized as follows. In Section~\ref{sec:calibration},  we give a general review of common calibration approaches for radio interferometric measurements. In Section~\ref{sec:corrcal}, we present \corrcal{} calibration scheme. In Section~\ref{sec:data} we apply our method to the 9-hour PAPER-64 data set. In Section~\ref{sec:results}, we present the results of the \corrcal{} calibration. The conclusion will be in the last section.

\section{A Review of Calibration Formalism}
\label{sec:calibration}
In radio interferometry, the correlated signal (visibility) from two independent receiving elements (antennas) is the sum of the system noise and the actual sky signal (or true sky visibility) multiplied by the complex gain parameters of the antennas. The gain parameters depend on the electromagnetic properties of the antennas, whereas the additive noise is mostly dominated by the thermal noise of the receiver system and the sky components.

Performing a calibration is to solve for the complex gain parameters to recover the true sky signal from the measured signal. The mathematical form of this statement is the measurement equation,
\begin{eqnarray}\label{eq:measurement_eq}
    d_{pq} = g_{p}\,g^{\ast}_{q}\,v_{pq}+\,n_{pq}. \label{eq:d-pq}
\end{eqnarray}
Here, $g_{p}$ and $g_{q}$ are the unknown complex gain parameters of antenna $p$ and antenna $q$, which depend on the frequency $f$. $v_{pq}$ is the true sky visibility corresponding to the baseline between antenna $p$ and antenna $q$, for which we ultimately want to recover from the measured visibility (observed data) $d_{pq}$. $n_{pq}$ is the system noise on this baseline, which is usually assumed to be Gaussian. The asterisk superscript $^{\ast}$ indicates the complex conjugate. All terms in~\refeq{eq:measurement_eq} are also implicitly time-dependent. The antenna gain $g$ can also be written as a complex exponential form with amplitude $a$ and phase $\phi$ for each of the antenna $p$ (or $q$),
\begin{eqnarray}
\label{eq:gain_parameter}
g_{p} = a_{p} \exp \left( \iu \phi_{p}  \right).
\end{eqnarray}
Note that we used ``$\iu$'' to indicate the imaginary unit $\sqrt{-1}$ throughout the paper.

For multiple baselines, the measurement equation can be written in a more compact form by using vector and matrix notations
\begin{eqnarray}\label{eq:measurement_eq_matrix}
    \d = \G \v + \n.
\end{eqnarray}
Here, $\d$ is a vector containing the measured visibilities of different baselines. $\v$ is vector of the corresponding true sky visibility. $\G$ is a diagonal gain matrix with $g_{p}g^{\ast}_{q}$ as its elements placed diagonally, and $\n$ is a vector containing noise of the corresponding baseline formed from antenna $p$ and $q$. For simplicity, we will omit the frequency (and time) dependency in the vector and matrix notations through the rest of the paper unless otherwise stated.

Most calibration software packages developed so far to solve for the gain parameter $\G$ can be categorized into either one of the two conventional approaches \emph{sky-based} calibration or \emph{redundant-baseline} calibration\footnote{It is worth noticing that~\citet{Liu2010} investigated the Taylor expansion of the sky to deal with direction-dependence or non-redundancy.}. In the sub-sections that follow, we present the underlying mathematical formalism of these two calibration approaches and discuss their inherent advantages and disadvantages. 

\subsection{Sky-based Calibration}
\label{sec:sky_based}
Sky-based calibration uses a model visibility that approximates the true sky visibility, $m_{pq} \approx v_{pq}$ to solve for the gain solutions. ~\refeq{eq:d-pq} can be written equivalently in the matrix form,
\begin{eqnarray}\label{eq:skybased_matrix}
    \d = \G \m + \n.
\end{eqnarray}
The model visibility is usually constructed from known sources and the telescope beam response. Solving for the gains $\G$ is then essentially equivalent to minimizing the $\chi^{2}$ function of the difference between the measured visibility $\d$ and the model visibility $\m$, 
\begin{eqnarray}\label{eq:skybase_chisq}
    \chi^{2} = (\d-\G \m)^{\dagger}
        \N^{-1}
        (\d - \G \m).
\end{eqnarray}
Here, $\N$ is an ensemble average noise matrix in which its diagonal elements are the noise variance $\sigma_{pq}^{2}$ on the baseline formed from antenna $p$ and antenna $q$,
\begin{eqnarray}\label{eq:N}
\label{eq:noise_matrix}
\N &=&  \langle\n\n^{\dagger} \rangle  \nonumber \\
   &=&  \sigma_{pq}^{2} \I,
\end{eqnarray}
where the angle brackets $\langle .. . \rangle $ denote ensemble average, and in the second equality in ~\refeq{eq:N} assumes that the noise matrix is diagonal, i.e., the noise on visibilities of different baselines are {\it uncorrelated}. $\I$ is the identity matrix, and $\dagger$ denotes a complex conjugate transpose. The measured visibility can be corrected (or calibrated) by using the gain solutions estimated from optimization of~\refeq{eq:skybase_chisq}. However, the precision of sky-based calibration is highly dependent on the sky models, which are imperfect due to missing sources or an inaccurate beam model. This imperfection can lead to calibration errors that introduce foreground contamination in the cosmological Fourier space for 21 cm signal~\citep{Beardsley2016, Barry2016, Ewall2017, Byrne2019}.

\subsection{Redundant Calibration}
\label{sec:redundunt_cal}

Redundant calibration relies on multiple copies of identical baselines in arrays with the regular layout, to relatively solve for both the gain parameter and the true sky visibility from an over-determined system of linear equations without the need for the sky model \citep{Wieringa92}. The measurement equation for redundant calibration replaces the true sky visibilities $v_{pq}$ with visibility terms that are constrained to be equal across redundant baselines $u_{\alpha}$, where $\alpha$ indexes the redundant baseline sets (redundant groups),
\begin{eqnarray}\label{eq:redundant}
    d_{pq}= g_{p}g^{\ast}_{q} u_{\alpha} + n_{pq}.
\end{eqnarray}
Since the number of measured visibility $d_{pq}$  are greater than the number of $u_{\alpha}$ that averaged over redundant group of the array and the gain parameter $g$, this equation becomes over-determined and can be solved, provided there are sufficient redundant baselines. The solutions for~\refeq{eq:redundant} can be then determined by minimizing the $\chi^{2}$ function of the form, 
\begin{eqnarray}\label{eq:redundant_chisq}
    \chi^{2} = (\d - \G \u_{\alpha})^{\dagger}
        \N^{-1} (\d - \G\u_{\alpha}).
\end{eqnarray}
In contrast to~\refeq{eq:skybase_chisq}, the redundant visibility parameter $\u_{\alpha}$ here is not pre-defined and will be solved iteratively alongside with the antenna gain factor $\G \equiv g_{p}g_{q}$. To minimise the $\chi^{2}$ function, a least square estimator is usually utilised. We direct the interested reader to the existing literature \citep[e.g.,][]{Liu2010,Zheng2014,Dillon2018,Dillon20} for detailed discussion of least square solver implementation. Minimizing ~\refeq{eq:redundant_chisq} gives us a power spectrum estimation
 \begin{equation}
 \label{eq:least_square_estimator}
    \hatx = 
     [\A^{\intercal}\N^{-1}\A]^{-1}
     \A^{\intercal} \N^{-1} \d.
 \end{equation}
Here, the matrix $\A$, which depends on the array configuration, maps the measured visibility $\d$ to the calibration parameters $\G$ and $\v$ for which we want to solve. The vector $\hatx$ contains the least-square estimates power spectrum estimates for per antenna gain parameter, and per-redundant group averaged visibility.~\refeq{eq:least_square_estimator} is solvable iteratively, using the fitted calibration parameter as the fiducial guesses for the next cycle of the fitting, until the solutions converge. The method is implemented in the \omnical{}\footnote{\url{https://github.com/jeffzhen/omnical.git}} software package~\citep{Zheng2014}, which was used to calibrate PAPER and the MWA Phase II observations~\citep{Parsons2014, Ali2015, Li_2018, Kolopanis2019}.

Due to the \emph{degeneracies} in the measurement equation,~\refeq{eq:redundant}, the least square estimator of \refeq{eq:redundant_chisq} will not converge to a unique solution.  Here, the term degeneracies refer to the linear combinations of gains and visibilities that redundant calibration cannot solve for, and which results in the $\chi^{2}$ function is  invariant across different choices of the degenerate parameters. In-depth discussions of degeneracy issues in redundant baseline calibration can be found in ~\citep{Liu2010,Zheng2014,Dillon2018, Byrne2019}.

In general,  redundant baseline calibration suffers from four types of degeneracies for array elements configured on the same plane. First, if one multiplies the amplitude of the  antenna gain by a constant factor $A$ and divides the sky parameter by $A^{2}$, the $\chi^{2}$ remains unchanged so that the overall amplitude, $A$,  of the instrument can not be solved. Second, moving the phase of the gain parameter by certain amount $\Delta$, i.e., $g_{p}=|g_{p}|e^{j\phi_{p}}\rightarrow|g_{p}|e^{j(\phi_{p}+\Delta)}$ does not affect the $\chi^{2}$ due to cancellation effect of $g_{p}\times g^{*}_{q}$ in~\refeq{eq:measurement_eq} so that the redundant baseline calibration still can not be set the overall phase degeneracy $\Delta$. The third and fourth  degeneracies are caused by phase gradients $\Delta_{x}$ and $\Delta_{y}$ along $x$ and $y$ direction, respectively, for a planar array. For example, if one shifts $g_{p}=|g_{p}|e^{j\phi_{p}}\rightarrow|g_{q}|e^{j(\phi_{q}+\Delta_{x}x_{q}+\Delta_{y}y_{q})}$ it does not affect the $\chi^{2}$ if it is complemented by shift in sky parameter $V_{pq}^{\rm{true}}=|V_{pq}^{\rm{true}}|e^{j\phi_{pq}}\rightarrow|V_{pq}^{\rm{true}}|e^{j(\phi_{pq}-\Delta_{x}b_{x}-\Delta_{y}b_{y})}$, where $x_{q}$ and $y_{q}$ are the coordinates for antenna $j$, and $b_{x}=x_{q}-x_{p}$ and $b_{y}=y_{q}-y_{p}$ are $x$ and $y$ coordinates  of a baseline vector $\textbf{b}$ connecting antenna $p$ and $q$, respectively. So, the phase gradients $\Delta_{x}$ and $\Delta_{y}$ are still degenerate in the relative calibration stage.  In a sense, this means that inclining  the whole array in either direction (x or y) is exactly equivalent to  moving the phase centre of the sources on the sky in opposite direction. Hence, these two effects can cancel each other and result in the phase gradient degeneracies, which in turn results in the sources appearing to be offset from the phase centre. 

To fix the degeneracy issues, the redundant calibration can be divided into two calibration steps. These are relative calibration and absolute calibration. First, relative calibration is performed using the algorithm that we have just described to solve the antenna gains and the true sky. However, these solutions are degenerate, and an overall amplitude, an overall phase, and phase gradient calibration parameters are remain unknown.  Then, absolute calibration is performed to set these degeneracies to a reference and obtain the final calibration solutions.  One particular approach for absolution calibration, as discussed in~\citet{Byrne2019}, is to map a pre-defined sky model of bright point sources to the degenerate parameters. However, setting degeneracies using the sky information in redundant calibration is difficult because incompleteness of the sky model can introduce frequency-dependent gain errors, which may lead to spectral structures in the otherwise smooth observations~\citep{Barry2016, Byrne2019}.

Thus, the fundamental shortcomings of redundant calibration are imperfection of array redundancy and sky-modeling error. During the absolute calibration stage, the sky-modeling error enters into the redundant calibration. Redundant calibration assumes that the same sky is seen by all baselines in the same redundant group, requiring these baselines to have the same physical length and orientation. This assumption also dictates that all antenna elements in the array have the same primary beam responses. In the real situation, both are impossible to achieve for the level of accuracy required. These non-redundancies will result in calibration errors that lead to more foreground contamination in the cosmological Fourier space for 21 cm signal ~\citep{Ewall2017,Orosz2018}. The powerful alternative is to include the relaxed-redundancy information and position of sources into the calibration formalism~\citep{S17}. 

\section{Correlation Calibration (\corrcal)}
\label{sec:corrcal}
\citet{S17} proposed a hybrid calibration scheme called \corrcal{}, which aims to further bridge the gap between sky-based and redundant calibrations by taking into account sky and array information into the calibration algorithm. The \corrcal{} framework relies on the assumption that sky information is statistically Gaussian. While the real sky is never exactly Gaussian, ~\citet{S17} argued that the assumption still provide a good approximation of the reality. This assumption allows \corrcal{} to relax the redundancy of the array and including the known sky information in its formalism through covariance based calculation of $\chi^{2}$. The covariance matrix thus forms  a \emph{statistical model} of the sky power spectrum, weighted by the instrumental response. We describe the underlying mathematical formalism of \corrcal{} in this section.

Given a vector of the measured visibilities $\d$ as is defined in~\refeq{eq:measurement_eq_matrix}, its covariance $\Sig1$ follows,
\begin{eqnarray}
\Sig1 = \langle  \d \d^{\dagger} \rangle.
\end{eqnarray}
Using the definition of $\d$ from~\refeq{eq:measurement_eq_matrix}, $\Sig1$ can be expanded as,
\begin{eqnarray}
\label{eq:visibility_correlation}
\Sig1 &=&  \langle (\G \v + \n)
    (\G \v + \n)^{\dagger} \rangle \nonumber \\
    &=&  \G\langle \v \v^{\dagger}\rangle\G^{\dagger}
    + \G\langle\v \n^{\dagger} \rangle
    + \langle \n \v^{\dagger}\rangle\G^{\dagger} 
    + \langle \n \n^{\dagger} \rangle.
\end{eqnarray}
By assuming that the instrument noise vector $\n$ does not correlate with the true sky visibility vector $\v$, the middle two terms drop out. The last term describes the instrument noise, which is generally assumed to be Gaussian and uncorrelated between baselines; therefore, it can be written as a noise matrix $\N$ as defined in~\refeq{eq:noise_matrix}. The first term describes the correlation of the sky multiplied by the gains. If we define the variance-covariance matrix of the true sky as
\begin{eqnarray}
\label{eq:covariance_matrix}
\C = \langle \v \v^{\dagger} \rangle,
\end{eqnarray}
~\refeq{eq:visibility_correlation} takes
\begin{eqnarray}
\label{eq:visibility_correlation_matrix}
\Sig1 = \G \C \G^{\dagger} + \N.
\end{eqnarray}
Under an assumption that the sky is random Gaussian distributed, we note that by virtue of the central limit theorem the multitude of random phenomena that produce the random character of the observed data $\d$,  implies their distributions are nearly Gaussian in general. Thus, the likelihood $\mathcal{L}$ of data $\d$ given their covariance matrix $\Sig1$ takes the standard form of a multi-dimensional Gaussian distribution,
\begin{eqnarray}
   \mathcal{L}(\d,\Sig1) \propto \exp\left[-\frac{1}{2}\d^{\dagger}\Sig1^{-1}\d\right].
\end{eqnarray}
 Using~\refeq{eq:visibility_correlation_matrix} in this equation, we obtain the likelihood  
\begin{eqnarray}\label{eq:corrcal_likelihood}
   \mathcal{L}(\d,\C|\G) \propto  \exp\left[-\frac{1}{2}\d^{\dagger}\left(\G\C\G^{\dagger}+\N\right)^{-1}\d\right].
\end{eqnarray}
Taking the logarithm of both sides of \refeq{eq:corrcal_likelihood}, we obtain the negative log-likelihood  function $\log\mathcal{L}$,
\begin{eqnarray}\label{eq:corrcal_log_likelihood}
    \log\mathcal{L}(\d,\C|\G) \propto
    -\frac{1}{2}\d^{\dagger}\left(\G\C\G^{\dagger}+\N\right)^{-1}\d .
\end{eqnarray}
Maximising the likelihood function $\mathcal{L}$ is equivalent to minimising the negative log-likelihood $\log\mathcal{L}$. The term on the right-hand side of~\refeq{eq:corrcal_log_likelihood} measures the square of the distances from predicted (or true) visibility to the measured visibility in the standard deviation unit.  Minimising this term against $\G$ is equal to maximising the likelihood of true visibility being observed.  The exponential term is thus simply equal to the $\chi^{2}$, given by
\begin{eqnarray}\label{eq:corrcal_chisq}
    \chi^{2}=\d^{\dagger}\left(\G\C\G^{\dagger}+\N\right)^{-1}\d.
\end{eqnarray}
The covariance based calculation of $\chi^{2}$ in this equation is the underlying mathematical formalism of \corrcal{}. We further add an additional term $\chi^{2}\rightarrow \chi^{2}+(\sum_{i}{\rm Im}(g_{i}))^2+
(\sum_{i}{\rm Re}(g_{i})-N_{\rm ant})^2$ as a regularisation factor to bound the gain solution. Here, the $\chi^{2}$ function does not explicitly depends on the sky parameter or written in terms of the redundant sets of visibility, but, still, the array and sky information are integrated into the covariance matrix $\C$ of the true sky visibility. 

Since we are modeling the data as being Gaussian distributed, a form of the $\chi^{2}$ function in ~\refeq{eq:corrcal_chisq} is different from the one defined in ~\refeq{eq:skybase_chisq} and ~\refeq{eq:redundant_chisq} which explicitly depends on the sky parameters $\m$ and $\u_{\alpha}$, respectively. However, the \corrcal{}  approach of keeping the array  and sky information in the $\C$ matrix allows one to reproduce an exact copy of the redundant calibration algorithm in a limiting case. For instance, all baselines within the perfectly redundant group see the same sky, so the effective noise $\mathbf{N}_{\rm eff}$ for this group is written as the sum of per-visibility diagonal noise variance matrix $\N$ and covariance between visibilities. That is,
\begin{eqnarray}
\mathbf{N}_{\rm eff}=\N+(a\mathbf{1})\otimes(a\mathbf{1})^{\dagger},
\end{eqnarray}
where $a$ is a parameter controlling the variance of the sky signal, $\mathbf{1}=[1,1,1,\dots]$ is a vector of ones, and an operator $\otimes$ denotes the outer product of vectors. Applying the Woodbury identity (see ~\refeq{eq:woodbury}) to $\mathbf{N}_{\rm eff}^{-1}$, and taking  $a\to\infty$, the $\chi^{2}$ becomes the same to one for redundant calibration algorithm.

Although complete information of the true sky can never be obtained, we can incorporate partial sky information into the covariance matrix by using known sky models. Thus, we further split the covariance matrix in ~\refeq{eq:corrcal_chisq} into two components as per suggestion in ~\cite{S17}, 
\begin{eqnarray}\label{eq:csr_matrices}
\C = \S + \R,
\end{eqnarray}
where the matrix $\S$ contains the covariance of the expected visibilities of known sky sources, whereas $\R$ is the covariance matrix of the visibilities within the redundant group that contains everything else that is not in $\S$, i.e. `the rest,' which is mostly the diffuse sky signal from our Galaxy's emission, weighted by the instrumental response. Note that~\refeq{eq:csr_matrices} assumes that the sky components in $\S$ and $\R$ are uncorrelated since most point sources are extragalactic. With this definition, ~\refeq{eq:corrcal_chisq} becomes
\begin{eqnarray}\label{eq:chisq_with_src}
\chi^{2}=\d^{\dagger}\left(\G\big(\S+\R)\G^{\dagger}+\N\right)^{-1}\d.
\end{eqnarray}
Note that the explicit formalisms of $\S$ and $\R$ will be presented in  Section~\ref{sec:cov_matrices} in details.

With the redundant group diffuse sky $\R$ and sources $\S$ expressed as vector outer products, ~\refeq{eq:csr_matrices} can be rewritten as
\begin{eqnarray}\label{eq:outer_product}
\C=\s\s^{\dagger}+\r\r^{\dagger},
\end{eqnarray}
where $\r$ and $\s$ are $N$-dimensional vectors for the diffuse sky and source components, respectively. We substitute ~\refeq{eq:outer_product} for $\C$ in ~\refeq{eq:chisq_with_src} to obtain a $\chi^{2}$ function 

\begin{eqnarray}
\chi^{2}=\d^{\dagger}\left(\N+\G\big(\s\s^{\dagger}+\r\r^{\dagger}\big)\G^{\dagger}\right)^{-1}\d.
\end{eqnarray}
Applying the gain matrix to the source and redundant vectors in this equation, we have
\begin{eqnarray}
\label{eq:modified_chisq} 
\chi^{2}=\d^{\dagger}\Big(\N+\hat{\s}\hat{\s}^{\dagger}+\hat{\r}\hat{\r}^{\dagger}\Big)^{-1}\d,
\end{eqnarray}
where $\hat{\s}=\G\s,\hat{\r}=\G\r$, so $\hat{\s}^{\dagger}=\s^{\dagger}\G^{\dagger}$ and $\hat{\r}^{\dagger}=\r^{\dagger}\G^{\dagger}$.

Inversion of the matrices inside the parenthesis term in ~\refeq{eq:modified_chisq} can be done with Woodbury inversion formula~\citep{Woodbury50}, which we write here in the special case of a Hermitian matrix if both $\B^{-1}$ and $\I+\v^{\dagger}\B^{-1}\v$ are invertible 

\begin{align}\label{eq:woodbury}
\Big(\B +\v\otimes\v^{\dagger}\Big)^{-1}=\B^{-1}
-\B^{-1}\v\Big(\I+\v^{\dagger}\B^{-1}\v\Big)^{-1}
\v^{\dagger}\B.
\end{align}
Where $\B$ and  the identity matrix $\I$ are square matrices with dimension $N\times N$, while $\v$ is $N$-dimensional vector.  
	

For the sake of computational tractability, the implementation of the Woodburry inversion to  ~\refeq{eq:modified_chisq} in \corrcal{}  follows the following steps:
\begin{itemize}
	\item First, the  Woodburry inversion can be applied to $\Gamma^{-1}$ where $\Gamma=\big(\N+\hat{\r}\hat{\r}^{\dagger}\big)$ for each redundant group separately, ignoring the source vectors $\s$. To apply Woodburry identity to $\Gamma^{-1}$, let $\B=\N$ and $\v=\r$. Then, the inverted form of  $\Gamma$ is kept separate for the each redundant group, and saved in a factored form by applying the Cholesky factorization of the matrix $\big[\I+\v^{\dagger}\B^{-1}\v\big]^{-1}$ in ~\refeq{eq:woodbury} for the redundant vectors $\r$.   
	\item Finally,  using the source vectors $\s$, and the already factored form of $\Gamma^{-1}$ from the first step the Woodburry inversion can be computed using $\big(\Gamma+\hat{\s}\hat{\s}^{\dagger}\big)^{-1}.$
\end{itemize}
\subsection{Determining the Sky and Noise Variance-Covariance Matrices}
\label{sec:cov_matrices}
To perform the \corrcal{}, we first need to estimate the
matrices $\R$, $\S$, and $\N$. Thus,  in this section, we will derive these variance-covariance matrices explicitly.  

\subsubsection{Diffuse Sky Component Matrix ($\R$)}
\label{sec:r_matrix}
For an instrument with a small field of view, the covariance matrices can be directly constructed from the visibility in the 2D UV plane. However, most 21-cm experiments, including PAPER, use receiving elements that have wide field-of-views. Thus, the curvature of the sky becomes important. To account for this effect and simplify the calculation, we will first project the visibility function into the spherical harmonic space before calculating the covariance. 

Given a vector $\u = \b / \lambda$ that expresses the baseline vector $\b$ between antennas $p$ and $q$ in wavelength $\lambda$, the measured cross-correlation signal from these antennas, or the visibility, ignoring noise term, is given by,
\begin{eqnarray}
v_{pq}(\u| f)
= \int \textrm{d}\Omega\, I(\ur, f) B(\ur, f),
\label{eq:visibility_eq}    
\end{eqnarray}
where 
\begin{eqnarray}
B(\ur,f) &=& A_{p}(\ur,f)A^{\ast}_{q}(\ur, f) \exp\left[ -2\iu\pi \u\cdot\ur \right]\nonumber\\
&=&|A_{p}(\ur,f)|^{2}\exp\left[ -2\iu\pi \u\cdot\ur \right].
\label{eq:beam_transfer_func}
\end{eqnarray}
In this equation,  $I(\ur, f)$ is the flux density measured over a solid angle $\Omega$ subtended by the observed region of the sky in the direction of a unit vector $\ur$. $A_{p}(\ur, f)$ and $A_{q}(\ur, f)$ are the primary beam of antenna $p$ and $q$, repectively.
Here, the squaring of the antenna beam term $A_{p}(\ur,f)$ in the second equality of \refeq{eq:beam_transfer_func} comes from the fact that we assume all antennas have the same primary beam response. The antenna position information is integrated into the exponential factor. Thus, the second line of the equation carries array information (position and beam response). 

Expanding the terms for the sky ($I(\ur, f)$) and array ($B(\ur, f)$) over the spherical harmonics, we have
\begin{eqnarray}\label{eq:sph_intensity}
I(\ur, f)=\sum_{\ell m}a_{\ell m}(f)Y_{\ell m}(\ur),
\end{eqnarray}
\begin{eqnarray}
B(\ur, f)=\sum_{\ell m}b_{\ell m}(f)Y_{\ell m}(\ur), 
\label{eq:sph_beam}
\end{eqnarray}
where $a_{\ell m}$  and $b_{\ell m}$ are the amplitudes of the flux density and the beam function for the spherical harmonics $Y_{\ell m}$ respectively. The angular number $\ell=0,1,2,3,\dots$, and the azimuthal number $m=-\ell, -\ell+1,\cdots, \ell-1, \ell$ follow standard convention, allowing $2\ell+1$ values of $m$ for each value of $\ell$. Using~\refeq{eq:sph_intensity} and~\refeq{eq:sph_beam}, the visibility in~\refeq{eq:visibility_eq} can be expressed in spherical harmonics as,
\begin{eqnarray}
\label{eq:spherical_harmonic_vis}
v_{pq}(f)&=& \int \textrm{d}\Omega\,
\sum_{\ell m}a^{\ast}_{\ell m}(f)Y^{\ast}_{\ell m}(\ur)
\sum_{\ell' m'}b_{\ell' m'}(f)Y_{\ell' m'}(\ur) \nonumber \\
&=& \sum_{\ell \ell' mm'}a^{\ast}_{\ell m}(f)b_{\ell' m'}(f)
\int\textrm{d}\Omega\, Y^{\ast}_{\ell m}(\ur)Y_{\ell' m'}(\ur) \nonumber \\
& =& \sum_{\ell m}a^{\ast}_{\ell m}(f)b_{\ell m}(f),
\end{eqnarray}
where we have used the orthogonality of spherical harmonics to turn the integral in the second line into a Kronecker delta function.~\refeq{eq:spherical_harmonic_vis} is the spherical harmonics version of the visibility equation in the absence of noise.

\begin{figure}
	\centering
	\includegraphics[width=\columnwidth]{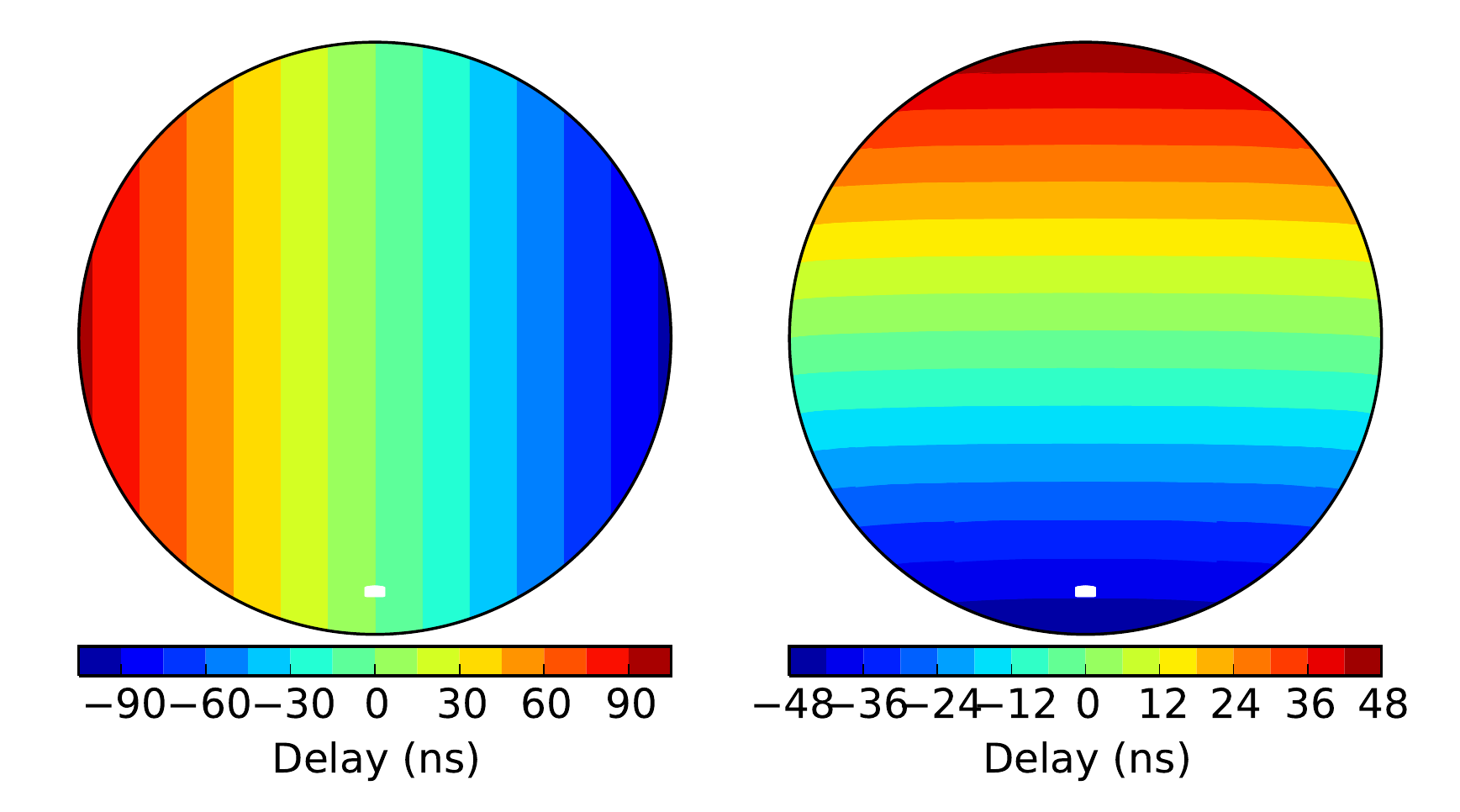}
	\caption{A delay term ($\b\cdot\hat{\mathbf{r}}/c$)  pattern in nanoseconds (ns) over PAPER latitude for baselines pointed along the East-West (left) and the North-South (right) directions. These baselines are calculated from the actual PAPER-64 antenna coordinates that shown in Fig.~\ref{fig:antpos}.}
	\label{fig:finge_pattern}
\end{figure}

\begin{figure}
	\centering
	\includegraphics[width=\columnwidth]{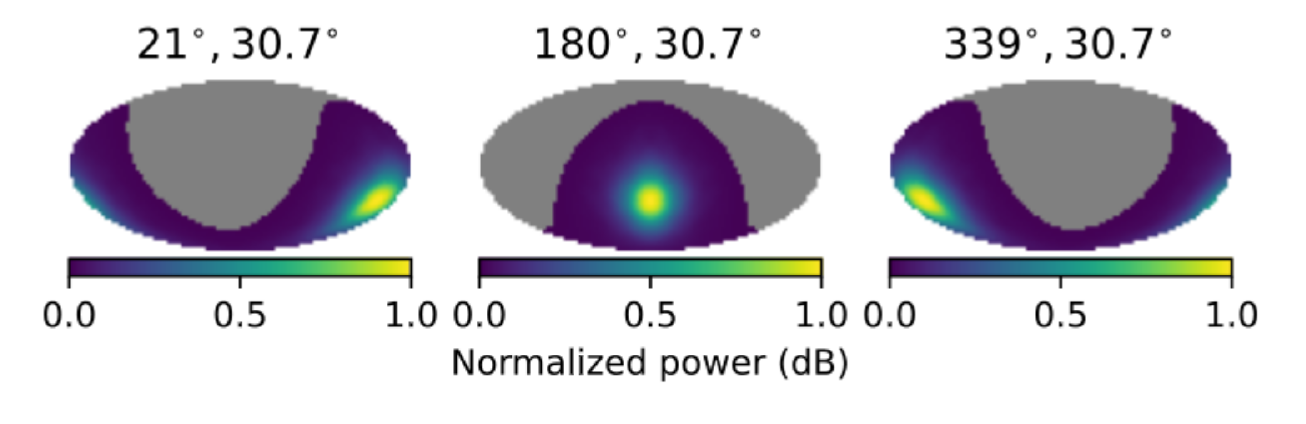}
	\caption{The AIPY simulated antenna beam pattern of PAPER-64 array at 150\,MHz at different longitudes, plotted for the PAPER latitude ($30.7^{\circ}$S) at Karoo desert, South Africa.}
	\label{fig:beam}
\end{figure}

We calculated the array function that is defined in \refeq{eq:beam_transfer_func} using the exponential argument ($\b\cdot\ur{}/c$) in~\refeq{eq:visibility_eq} and the PAPER-64 beam model depicted in Fig.~\ref{fig:beam}. The simulated exponential argument is shown in Fig.~\ref{fig:finge_pattern} for the PAPER-64 baselines oriented along the East-West and the North-South directions. To generate the beam model of the PAPER-64 array within the frequency range of 120\,MHz to 168\,MHz,  we use the  Astronomical Interferometry in Python (AIPY)\footnote{\url{https://github.com/HERA-Team/aipy.git}}, a software package primarily developed for PAPER/HERA data analysis. Fig.~\ref{fig:beam} shows the AIPY simulated PAPER-64 beam model projected on the HEALPIx\footnote{\url{ https://github.com/healpy/healpy.git}} coordinates at three arbitrary right accessions for frequency of 150\,MHz. We will use this beam model for all calculations of the covariance matrices.

\begin{figure}
	\centering
	\includegraphics[width=\columnwidth]{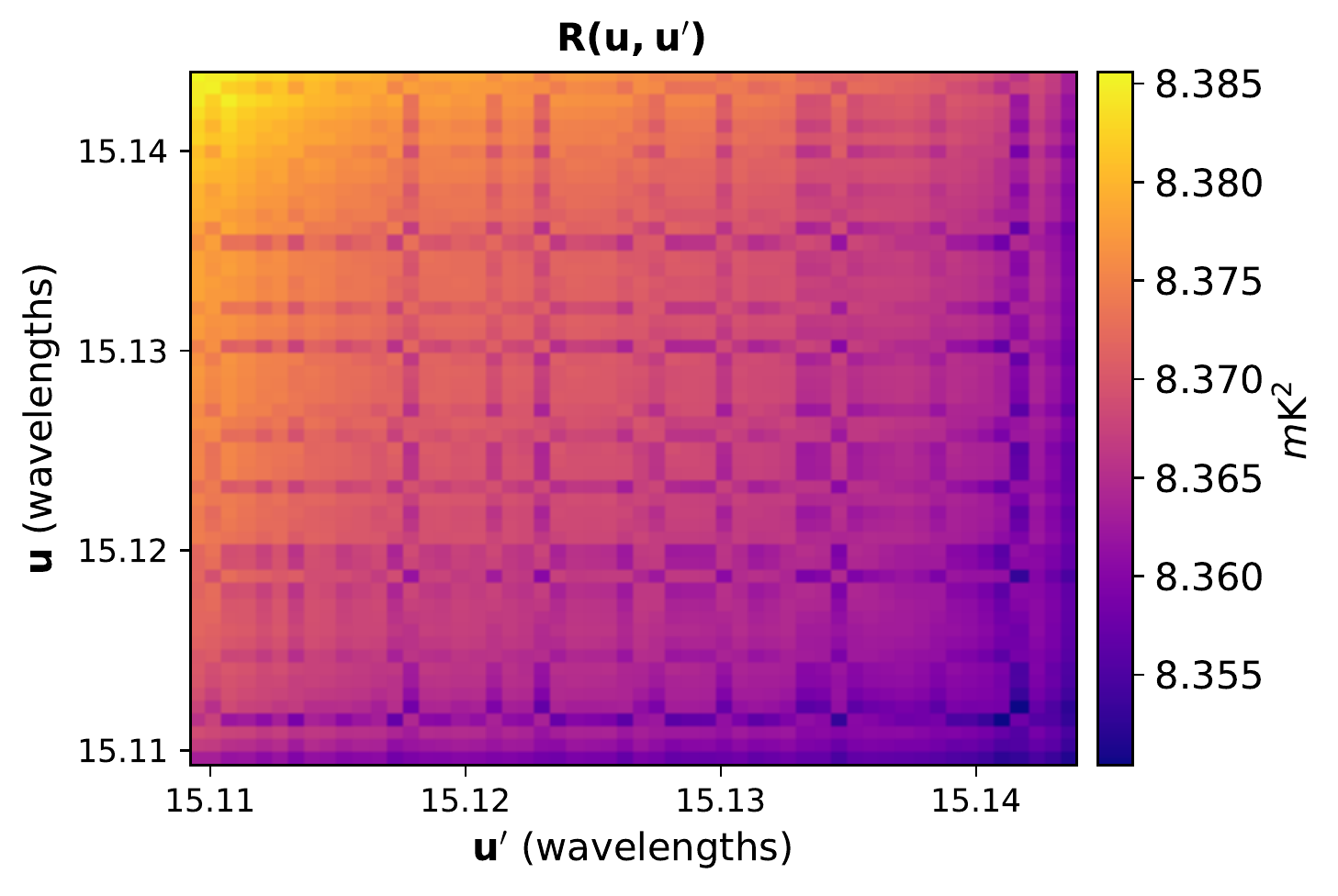}
	\caption{Covariance matrix between baselines obtained from the diffuse sky component at $150\,{\rm MHz}$ for a redundant group containing about $30$\,m length baselines. The axes of this figure spanning the true measured variation in the baseline lengths for the baselines in the 30m group.   The entries of the  covariance are not precisely equal as displayed in the figure. The covariance between baselines drops off as baselines become less redundant.}  
	\label{fig:covariance_matrix}
\end{figure}

Using the expected  visibility function defined in~\refeq{eq:spherical_harmonic_vis}, in this section we construct the diffuse sky covariance matrix $\R$.  While determining this matrix, all degrees of freedom including frequencies and baselines must be taken into accounts. We estimate this matrix from the relation  
\begin{eqnarray}
\label{eq:harmonic_com_mat}
\R_{({\u}f);({\u}^{\prime}f^{\prime})} &=&\left\langle v^{\rm diff}(\u,f)v^{{\rm diff},*}(\u^{\prime},f^{\prime})\right\rangle \nonumber \\
&= & \left\langle \sum_{\ell m}a^{{\rm diff},\ast}_{\ell m}(f)b^{{\rm diff}}_{\ell m}({\u},f) \right. \nonumber \\
&\times & \left. \sum_{\ell^{\prime}\mathrm{m}^{\prime}}a^{{\rm diff}}_{\ell^{\prime}\mathrm{m}^{\prime}}(f^{\prime})b^{{\rm diff},\ast}_{\ell^{\prime}\mathrm{m}^{\prime}}({\u}^{\prime},f^{\prime}))\right\rangle \nonumber \\
&= & \sum_{\ell m}\sum_{\ell^{\prime}\mathrm{m}^{\prime}}
b^{{\rm diff}}_{\ell m}({\u},f)\left\langle a^{{\rm diff},\ast}_{\ell m}(f)a^{{\rm diff}}_{\ell^{\prime}\mathrm{m}^{\prime}}(f^{\prime})\right\rangle \nonumber \\
&\times &
b^{{\rm diff},\ast}_{\ell^{\prime}\mathrm{m}^{\prime}}({\u}^{\prime},f^{\prime}),
\end{eqnarray}
where $^{\rm "diff"}$ is to denote $ \v_{pq}(f)$ is coming from the diffuse sky component and the angle bracket $\langle ...\rangle$ shows the averaging over an ensemble of realizations of the fluctuation. With the assumption that the sky is statistically isotropic Gaussian random field, the covariance for sky flux density spherical harmonics coefficients becomes 
\begin{eqnarray}
\left\langle a^{{\rm diff}}_{\ell m}(f)a_{\ell^{\prime}\mathrm{m}^{\prime}}^{{\rm diff},*}(f^{\prime})\right\rangle&=\left\langle|a_{\ell m}(f,f^{\prime})|^{2}\right\rangle\delta_{\ell\ell^{\prime}}\delta_{\mathrm{mm}^{\prime}} \nonumber  \\
&=C^{{\rm diff}}_{\ell}(f,f^{\prime})\delta_{\ell\ell^{\prime}}\delta_{mm^{\prime}}, \label{eq:aa-avg}
\end{eqnarray}
where $C^{{\rm diff}}_{\ell}(f,f^{\prime})=\langle|a^{{\rm diff}}_{\ell m}(f,f^{\prime})|^{2}\rangle$ is the the diffuse sky angular power spectrum, which only depends on multipole number $\ell$ that corresponds to the angular scale $\theta=180^{\circ}/\ell$, and $\delta$ is the Kronecker delta-function. This is because the assumption of isotropy ensures that $\langle|a^{{\rm diff}}_{\ell m}(f,f^{\prime})|^{2}\rangle$ is a function of only $\ell$, not $m$. Therefore, there is no correlation in the $m$ modes. Thus, the covariance matrix (\refeq{eq:harmonic_com_mat}) can be simplified to 
\begin{eqnarray}
\R=\sum_{\ell m}b^{{\rm diff}}_{\ell m}({\u},f)b^{{\rm diff},*}_{\ell m}({\u}^{\prime},f^{\prime})C^{{\rm diff}}_{\ell}(f,f^{\prime}).\label{eq:model_R}
\end{eqnarray}
Equation~(\ref{eq:model_R}) is a general expression for the covariance matrix corresponding to the diffuse sky. It depends on the baseline and antenna beam model through a function defined in ~\refeq{eq:beam_transfer_func}. The measured antenna positions (instead of assuming they are perfectly redundant) are integrated into this equation. Hence, through this expression, the covariance-based calculation of $\chi^{2}$ in~\refeq{eq:chisq_with_src} carries array information during optimization processes. Fig ~\ref{fig:covariance_matrix} displays the covariance matrix between visibilities at 150\,MHz in a given redundant group of baselines whose length about $30$\,m.

In~\refeq{eq:model_R}, however, the term $C^{{\rm diff}}_{\ell}(f,f^{\prime})$ is not known. The measured data covariance matrix should  be constructed from the same redundant group that $\R$ had formed to estimate $C^{{\rm diff}}_{\ell}(f,f^{\prime})$. The data covariance matrix can be determined from
\begin{eqnarray}\label{eq:18}
\D^{\alpha}
=\langle\d^{\alpha}({\u},f)\d^{\alpha\dagger}({\u}^{\prime},f^{\prime})\rangle,
\end{eqnarray}
where $\D^{\alpha}$ is a covariance matrix of measured visibility data $\d^{\alpha}$ from the redundant group $\alpha$. To determine the $C^{\rm diff}_{\ell}(f,f^{\prime})$, thus, we equate the covariance matrix $\D^{\alpha}$ with the diffuse sky  matrix $\R$ from the same redundant group $\alpha$ in~\refeq{eq:model_R}. Note that the diffuse galactic emissions dominate visibilities from shorter baselines,  while visibilities measured by longer baselines are unlikely to be dominated by diffuse emissions, so we only equate $\D$ with $\R$~(i.e., $\R^{\alpha}\approx\D^{\alpha}$) if both are computed from the same redundant group $\alpha$.  This assumption may avoid mode-mixing between baselines with different lengths. It follows that, 
\begin{eqnarray}\label{eq:sky}
C^{{\rm diff}}_{\ell}(f,f^{\prime})\approx\frac{\D^{\alpha}}{\sum_{\ell m}b^{{\rm diff}}_{\ell m}({\u},f)b^{{\rm diff}}_{\ell m}({\u}^{\prime},f^{\prime})}. 
\end{eqnarray}

\begin{figure*}
	\centering
	\includegraphics[width=16cm]{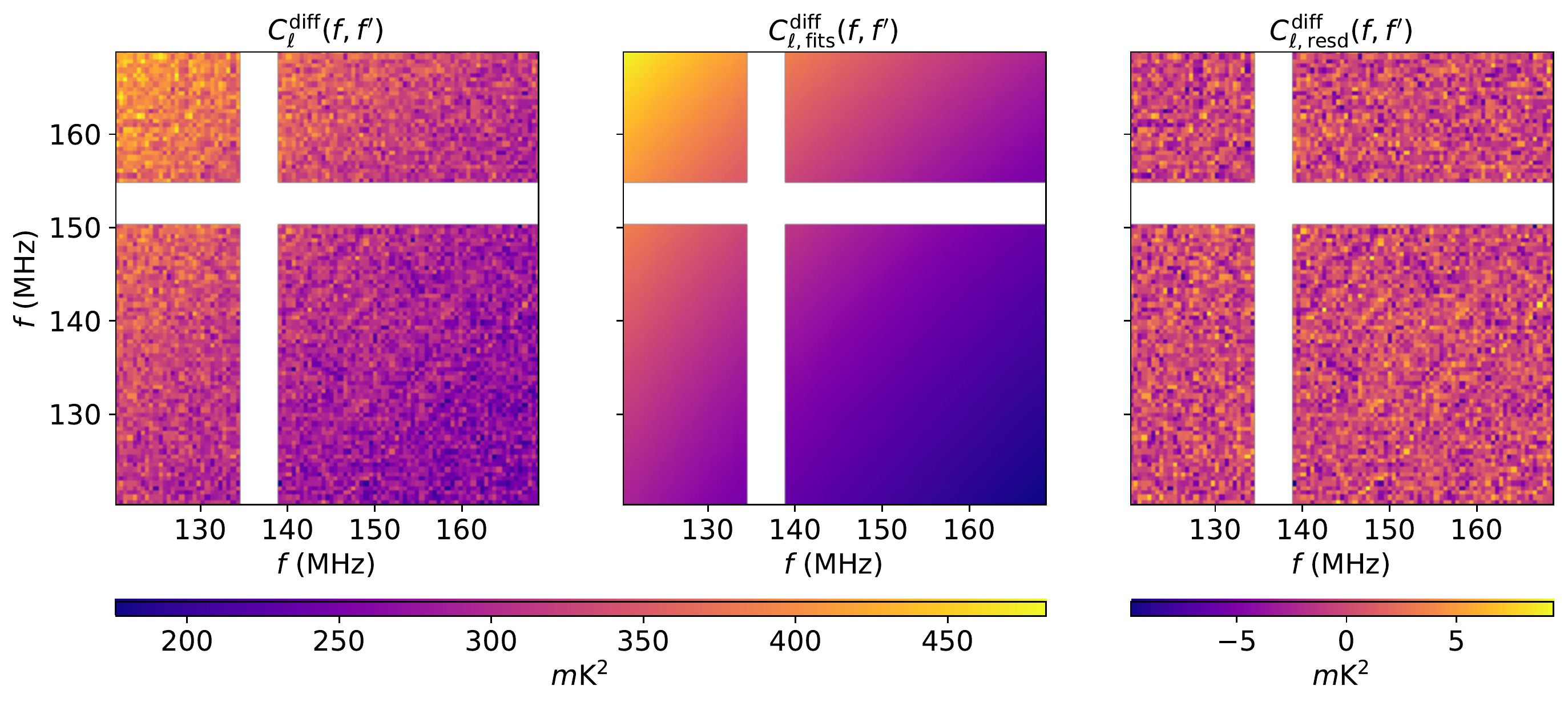}
	\caption{{\it Left}--Cross-frequency sky power spectrum $C_{\ell}^{{\rm diff}}(f,f^{\prime})$ obtained from~\refeq{eq:sky}.  {\it Middle}--the fitted power spectrum $C_{\ell,{\rm fits}}^{{\rm diff}}(f,f^{\prime})$. {\it Right}--the residuals $C^{\rm diff}_{\ell,\rm resd}(f,f^{\prime})=C_{\ell}^{{\rm diff}}(f,f^{\prime})-C_{\ell,{\rm fits}}^{{\rm diff}}(f,f^{\prime})$. In all plot we assume the $\ell$-dependence of $C^{\rm diff}_{\ell}(f,f')$ are the same. The whiteout regions are corresponding to the flagged frequency channels from the data.}
	\label{fig:fitted_cov} 
\end{figure*}

To pull $C^{\rm diff}_\ell(f,f^{\prime})$ out of the summation sign in~\refeq{eq:model_R}, we assume that for a given redundant block $\alpha$ the sky is slowly changing function of $\ell\approx 2\pi|\u|$. In other words, this means that the angular power spectrum of the sky does not evolve significantly over the baseline length within the redundant group.

Since sky power spectrum is real we discard the complex part of $C^{{\rm diff}}_{\ell}(f,f^{\prime})$ in \refeq{eq:sky}.  Then, we implement the power-law curve fitting into~\refeq{eq:sky} to model the angular cross-frequency power spectrum statistically. We make explicit assumptions about the functional form of the fitting function based on the fact that the power spectrum of the  diffuse emission follows a simple power-law with frequency~\citep{Santos2005}. We implement the {\tt Scipy} optimization routine for curve fitting to fit data with the pre-specified function. With the fitting parameters $A$ (amplitude in ${\rm mK^{2}}$) and the spectral index $\alpha$, the fitted power spectrum takes
\begin{equation}\label{eq:fits}
C^{\rm diff}_{{\rm fits}}(f,f')=A \left(\frac{ff'}{f^{2}_{0}}\right)^{-\alpha},
\end{equation}
where the fitted parameters $A=249.69\,{\rm mK^{2}}$ and $\alpha=1.5$, and the $f_{0}=150\,{\rm MHz}$ is the reference frequency. Since the power spectrum of the diffuse emission does not evolve significantly over baselines in the quasi-redundant group,  we assume that the fitted power spectrum is not a function of $\ell$.

Then the fitted angular power spectrum $C^{{\rm diff}}_{{\rm fits}}(f,f^{\prime})$ (see Fig.~\ref{fig:fitted_cov}) for that particular quasi-redundant group will be inserted in place of $C^{{\rm diff}}_{\ell}(f,f^{\prime})$  in~\refeq{eq:model_R} to estimate the model covariance matrix of the group. Likewise, one needs to derive the best-fit cross-frequency power spectrum statistically from the respective quasi-redundant sets to estimate the diffuse sky covariance matrix in each redundant group. In terms of the  fitted  $C^{{\rm diff}}_{{\rm fits}}(f,f^{\prime})$ parameter,  the diffuse sky covariance matrix $\R$ finally takes
\begin{eqnarray}
\label{eq:19}
\R=C^{\rm diff}_{{\rm fits}}(f,f^{\prime})\sum_{\ell}b^{{\rm diff}}_{\ell}({\u},f)b^{{\rm diff},*}_{\ell}({\u}^{\prime},f^{\prime}).
\end{eqnarray}

We use~\refeq{eq:19} to represent the diffuse sky component for  $\chi^{2}$ optimization process in~\refeq{eq:corrcal_chisq}.  The dimension of the covariance matrix $\R$ depends on the number of baseline $N_{\mathbf{\mathrm u}}$ and number of frequency channels $N_{f}$, i.e. its dimension is equal to $N_{\mathbf{\mathrm u}}N_{f}\times N_{\mathbf{\mathrm u}}N_{f}$.

\subsubsection{Point Source Component Matrix ($\S$)}
\label{sec:s_matrix}
The partial information about the point sources on the sky is one of the components the \corrcal{} calibration scheme employs to estimates the best likelihood antenna gain parameters.  Simulation in~\citet{S17} showed that the inclusion of the bright sources information in \corrcal{} has a significant effect in reducing both amplitude and phase calibration error.  However, the phase error has reduced remarkably (about a factor of 5 better than non-source informed correlation calibration) when the correlation information of point sources has implemented in~\citet{S17}, which substantially improves the quality of phase calibration. 

In this study, we use the published astronomical catalogues in~\citet{Jacobs2013} to  model the known sources. Using sources information in this catalogue, we form the matrix $\S$ that carries the sources information. 

The source position is implicit in time and usually expressed in terms of right ascension (RA) and declination (Dec) of catalogue source position. The time dependence of sources position vector $\ur_{i}$ thus comes from source motion relative to the observer meridian, which best visualised using the Hour-Angle (HA) and the observer Local Sidereal Time (LST). Because HA defines the amount of time since the source transited the observer meridian, hence, it tells how distant a source is from the meridian.  Based on the HA values (HA=LST-RA) of each source from the catalogue, we use sources that are above the horizon during observations time to create $\S$. We then project each source position onto the sphere. Then, we predict the visibilities in the spherical sky from a set of point sources with known positions.
Using the source statistics in this catalogue, we calculate the predicted visibility with the help of ~\refeq{eq:beam_transfer_func} for known sources. That is, suppose the point source profile is a Dirac-delta function on the sky $\ur_{i}$ ($i=1,2,...,N_{\rm ps}$, $N_{\rm ps}$ is the total number of point sources), then the specific intensity (\refeq{eq:visibility_eq}) can be written as

\begin{eqnarray}
I(\ur, f) =\sum^{N_{\rm ps}}_{i=1}I(\ur_{i},f)\delta_{\rm 2D}\left(\ur,\ur_{i} \right),
\end{eqnarray}
where $\delta_{\rm 2D}\left(\ur,\ur_{i} \right)$ is a two-dimensional Dirac delta function. Substituting this function into~\refeq{eq:beam_transfer_func}, we obtain the visibilities vector for point sources
\begin{eqnarray}
\label{eq:srv_vis_eq}
\v^{\rm ps}(\u,f)=\sum^{N_{\rm ps}}_{i=1}I(\ur_{i},f)\left|A_{pq}(\ur_{i},f) \right|^{2}{\rm e}^{-{\rm i}2\pi \u \cdot \ur_{i}}. \label{eq:visibility-ps}
\end{eqnarray}
To simulate beam function for sources, we use the known position on the sky from the established catalogue in~\citet{Jacobs2013} and the PAPER-64 simulated beam model that displayed in Fig.~\ref{fig:beam}. Taking the outer product betweenthe visibility  vector $\v^{\rm ps}(\u,f)$ and its complex conjugate transpose $\v^{\rm ps,\dagger}(\u,f)$, we form the covariance information for the discrete sources, 
\begin{eqnarray}\label{eq:ps_cov_mat}
\S=\v^{\rm ps}(\u,f)\v^{\rm ps,\dagger}(\u,f).
\end{eqnarray}
Thus, we will use this source information covariance matrix  to calculate the $\chi^{2}$ function in ~\refeq{eq:chisq_with_src}.
\subsubsection{Noise Matrix ($\N$)}
\label{sec:noise_matrix}
In this work, we estimate the per-diagonal noise variance matrix $\N$  from the \textit{radiometer equation} 
\begin{eqnarray}\label{eq:noise_var}
\sigma_{\rm rms}=\frac{T_{\rm sys}}{\sqrt{2\Delta f \tau}},
\end{eqnarray}
where $\Delta f=97\,{\rm KHz}$ is the frequency band width in KHz, $\tau=49.2$~s is the integration time. The system temperature $T_{\mathrm{sys}}$ from the relation defined in eq. (23) of \cite{Cheng2018} is,
\begin{equation}\label{eq:sys_temp}
T_{\rm sys}=180\,{\rm K}\left(\frac{f}{0.18\,{\rm GHz}}\right)^{-2.55}+T_{\rm rcvr}
\end{equation}
where $f$ is frequency, and $T_{\rm rcvr}$ is a receiver temperature which equal to 144 K. 
\begin{figure}
	\centering
	\includegraphics[width=8cm]{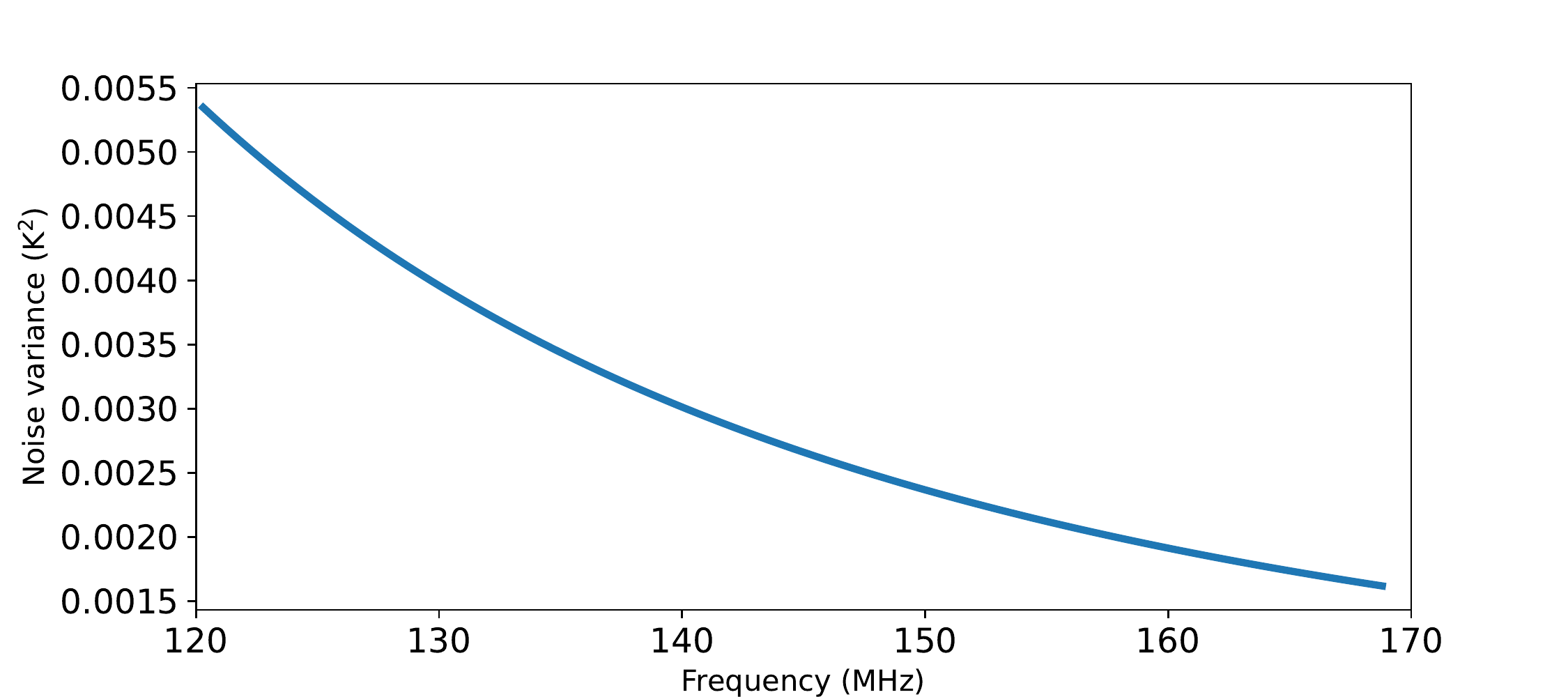}
	\caption{Noise variance estimated from the system temperature of PAPER.}
	\label{fig:noise_variance}
\end{figure}
\subsection{Eigendecomposition of sky signal covariance matrix}
The amplitude of $\R$ in ~\refeq{eq:19} for a perfectly redundant group containing two baselines is equal to a parameter controlling the variance of the sky, $C_{\ell}$,  times a square matrix of ones. That is,
\begin{eqnarray}\label{eq:example_r}
\R=C_{\ell}^{2}
\begin{bmatrix}
1 & 1  \\
1 & 1 \\
\end{bmatrix}\qquad ,
\end{eqnarray}

However, for a quasi-redundant group,  all the off-diagonal elements of the correlation matrix $\R$ are slightly different from one. As shown in Fig.~\ref{fig:covariance_matrix},  all the elements of the covariance matrix are not precisely equal. It shows that  the correlation between baselines decreases as they become less redundant. In that case, we approximate $\R$  from its eigendecomposition in our analysis following~\citet{S17}. 
\begin{figure}
	\centering
	\includegraphics[width=\columnwidth]{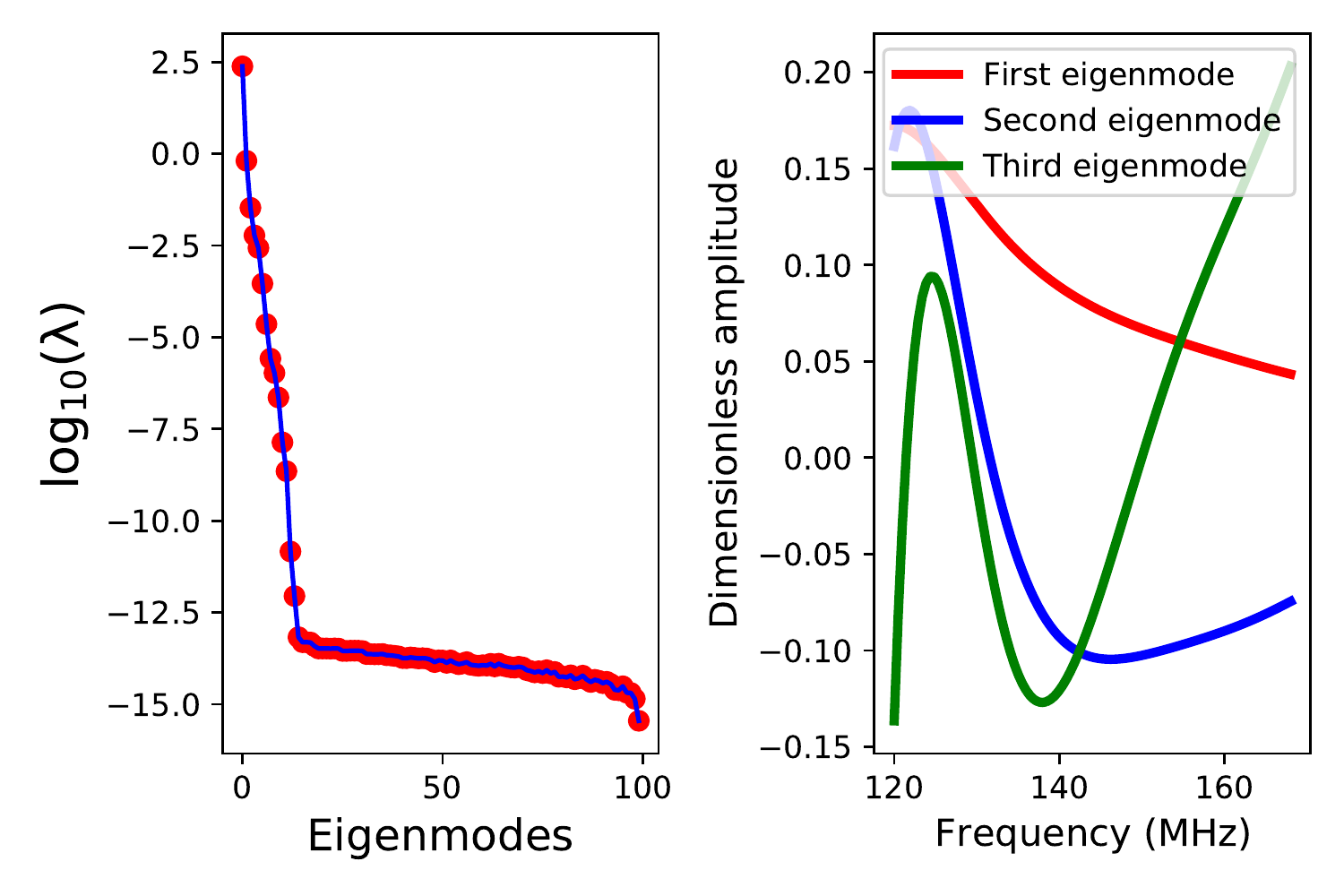}
	\caption{{\it Left}--Eigenvalues of cross-frequency covariance matrix described in~\refeq{eq:19}. The vertical axis represents the $\log_{10}$ eigenvalues, while the horizontal axis labels the corresponding principal component index. {\it Right}--The first three eigenmodes for $\R$ corresponding to the first three eigenvalues for a frequency range from 120 to 168 MHz.}
	\label{fig:eigen_val}
\end{figure}
If each redundant group in the array is not represented by few eigenmodes that are much less than the visibilities in the group, it shows that there is no sufficient redundancy in the array and therefore any calibration scheme that relies on the array redundancy may not give the accurate results. In the eigen-decomposition analysis, few modes of the principal components accounted for most of the variances in the original data. Thus, those eigenmodes of $\R$ corresponding to sufficiently large eigenvalues could replace its original data with very little loss of information. In the left panel of Fig.~\ref{fig:eigen_val}, we show the eigenvalues of cross-frequency covariance matrix described in~\refeq{eq:19}. The vertical axis represents the $\log_{10}$ eigenvalues, while the horizontal axis labels the corresponding principal component numbers. The sharp fall-off in the eigenvalues against the eigenmode index suggests that, by measuring a few modes of an eigenvalue one can account for most of the variation in the original data that forms $\R$. In addition, all the eigenvalues are greater than zero is because $\R$ is a positive-definite covariance matrix. In the right panel of Fig.~\ref{fig:eigen_val}, we depict the first three eigenmodes for $\R$ corresponding to the first three eigenvalues for a frequency range from 120 to 168 MHz. Specially, the first and the second eigenmodes are more important to explain the diffuse sky covariance between frequency channel because they are a relatively smooth function of frequency.

We keep eigenmodes with amplitude more than $10^{-5}$ times the largest. This  allows the keeping of $\sim$2-3 complex eigenmodes per quasi-redundant block. These few modes could represent most of the  variations in a given redundant group $\alpha$. For instance, the proportion of variance related  to the eigenvalues $\lambda_{1}$ and $\lambda_{2}$  can be calculated from
\begin{eqnarray}\label{eq:Eig_variance}
\left(\frac{\lambda_{1}+\lambda_{2}}{\lambda_{1}+\lambda_{2}+\cdots +\lambda_{p}}\right)\times 100,
\end{eqnarray}
where $p$ is the total eigenmodes. If the value of~\refeq{eq:Eig_variance} is greater than the threshold percentage $99\%$, we use eigenmodes corresponding to eigenvalues $\lambda_{1}$ and $\lambda_{2}$ to create a matrix that carries information about weights contributed by the variables in the original data to these eigenmodes. Variables with the highest correlation (in absolute value) with a principal component receive the highest weight in that component. This vector can be calculated from  the statistical relation
\begin{eqnarray}\label{pca_whights}
\p=\mathbf{e}_{i}\sqrt{\lambda_{i}},
\end{eqnarray}
where $(\mathbf{e}_{i},\lambda_{i})$ is the eigenvector-eigenvalue pairs for $\R$, and the subscript $i$ is the principal component index.  Using the vector outer product of $\p$ thus we estimate a  matrix that represents the diffuse sky covariance, given by
\begin{eqnarray}\label{eq:corr_coeff}
\R_{\rm PCA}=\p\otimes\p^{\dagger},
\end{eqnarray}
where $\R_{\rm PCA}$ is the diffuse sky covariance matrix calculated from the principal component analysis (PCA) of $\R$ for a given redundant block $\alpha$. The dimension of $\R_{\rm PCA}$ would significantly be reduced if very few eigenmodes could represent most of the variation in the original data. For instance, if we keep only two eigenmodes of $\R$ for a given redundant group containing four baselines, a complex vector $\p$ could be
\begin{eqnarray}\label{eq:pvector}
\p=[p_{1},p_{2},0,0],
\end{eqnarray}
where $p_{1}$ and $p_{2}$ are complex elements of a vector $\p$ that retained from eigenmode analysis. The outer product of this vector by its complex conjugate transpose takes	
\begin{eqnarray}\label{eq:matrix}
\p\otimes\p^{\dagger}=\begin{bmatrix} 
	p_{1}p^{*}_{1} & p_{1}p^{*}_{2} & 0&0 \\
	p_{2}p^{*}_{1} & p_{2}p^{*}_{2} & 0&0\\
	0 & 0 & 0&0 \\
	0 & 0 & 0&0\\
	\end{bmatrix},
\quad
\end{eqnarray}
here, the operator $^*$ denotes a complex conjugate. Adding the noise to the above matrix, we have
\begin{eqnarray}\label{eq:noise_plus_pp}
\Gamma=\N+\p\otimes\p^{\dagger}.
\end{eqnarray}

All off-diagonal elements of the noise variance matrix $\N$, which has the same size as $\p\otimes\p^{\dagger}$, are zero. Thus, $\Gamma$ has  only a relatively small number of nonzero elements and it can be considered to be sparse because most of its elements are zero. Therefore, it is wasteful to use the general dense matrix algebraic methods to inverting $\Gamma$ since most of the $O(N^{3})$ operations devoted to inverting the matrix involve zero operands. Hence, in terms of memory storage and computational efficiency, performing the sparse matrices operation that traversing only nonzero elements have significant advantages over their dense matrix counterparts. Taking advantage of the sparse structure of the correlation matrix $\Gamma$, \corrcal{} utilize the sparse matrix operations to increases its computational performance in a large dataset.

To implement the sparse matrix computations in \corrcal{}, per-redundant group  correlation matrices $\Gamma$ are stored as a sub-matrices  on the diagonal of the square block-diagonal matrix. These sub-matrices will be square, and their dimensions vary depending on the number of baselines in the redundant group. Since the off-diagonal elements of the block-diagonal matrix are zero, and there can be many sparse sub-matrices stored as its diagonal elements, the overall matrix must be sparse. It can be easily compressed and thus requires significantly less storage which also enhances the computational speed.

\subsection{Implementation of \corrcal{}}
\label{sec:applying_corrcal}
The crucial steps to perform \corrcal{} are grouping data according to their corresponding redundant baseline and determining the matrices $\R$, $\S$, and $\N$ to each of these redundant groups.  Then, one needs to set up the sparse matrix representations for these matrices, determine the $\chi^{2}$ function and its gradient information using the \corrcal{} routines. Utilizing these ingredients as inputs, \corrcal{}  finally adopts the \scipy\footnote{\url{https://github.com/scipy/scipy.git}}built-in optimization algorithm to fits antenna gains.  We summarise the procedures of implementation of \corrcal{} as follows:
\begin{itemize} 
\item First, we need to create the sparse matrices representing $\R$, $\S$, and $\N$ set up to each redundant group using the \corrcal{} routine, \sparse. The following arguments must be supplied to this routine to create sparse matrices:

\begin{enumerate}
	\item The real-imaginary separated per-redundant group diffuse sky vector $\p$. In this vector, we make all its entries equal to zero except those entries  corresponding to eigenmodes that we kept per redundant block (see ~\refeq{eq:pvector}).
	\item The real-imaginary separated vector $\s$ that carries information about the known point sources.
	\item  Per-diagonal noise variance matrix $\N$.
	\item A vector that contains indices setting off the redundant group.
\end{enumerate}

\item Second, the $\chi^{2}$ function and its gradient information should be determined using the \corrcal{} subroutines. The arguments passed to  these subroutines in order to calculate the $\chi^{2}$ function and the gradient information  of $\chi^{2}$ are:
\begin{enumerate}
	\item A vector containing the real-imaginary separated initial guess for antenna gains.
	\item Per-redundant group real-imaginary separated measured visibility data vector with imaginaries follow their respective reals (e.g., see vectorised visibility in~\refeq{eq:vec_vis2}).
	\item The sparse matrix representations for $\R$, $\S$ and $\N$ from the first step.
	\item  A vector containing per-visibility antenna indices. 
\end{enumerate}

\item Finally, using the $\chi^{2}$, its gradient information, and a vector of the initial guess for antenna gains that scaled by a large number as an input, \corrcal{} adopts the \scipy{} built-in nonlinear conjugate gradient algorithm,  \optimize{}, to solve for antenna gains.
\end{itemize} 
\subsubsection{Antenna gain solutions as a function of frequency-bandpass solution}
\label{sec:gain_solutions}
For the EoR studies, highly precise frequency-dependent bandpass calibration plays a crucial role.  If one's calibration solution has a noise-like structure along the frequency axis, it may introduce fine-frequency structure to the otherwise smooth sky observations even if the instrument's response is the slow function of observing frequencies. It poses the serious challenges toward efforts to isolate spectrally structured weak 21-cm signal from spectrally-smooth strong foreground signal. Several strategies have proposed for ensuring smooth calibration solutions as a function of frequency. For instance, simulation in~\citet{Ewall2017} and~\citet{Orosz2018} has shown that down-weighting measurements from longer baselines in a redundant baseline calibration solution would eliminate the migration of fine spectral structure from longer baselines to the shorter baselines regimes. This effect is because the longer baselines have more chromatic nature than the shorter ones as shown in~\citet{Morales2012}.  Other strategies implement fitting the calibration solutions with smooth function in frequency for a bandpass calibration.  

\begin{figure}
	\centering
	\includegraphics[width=\columnwidth]{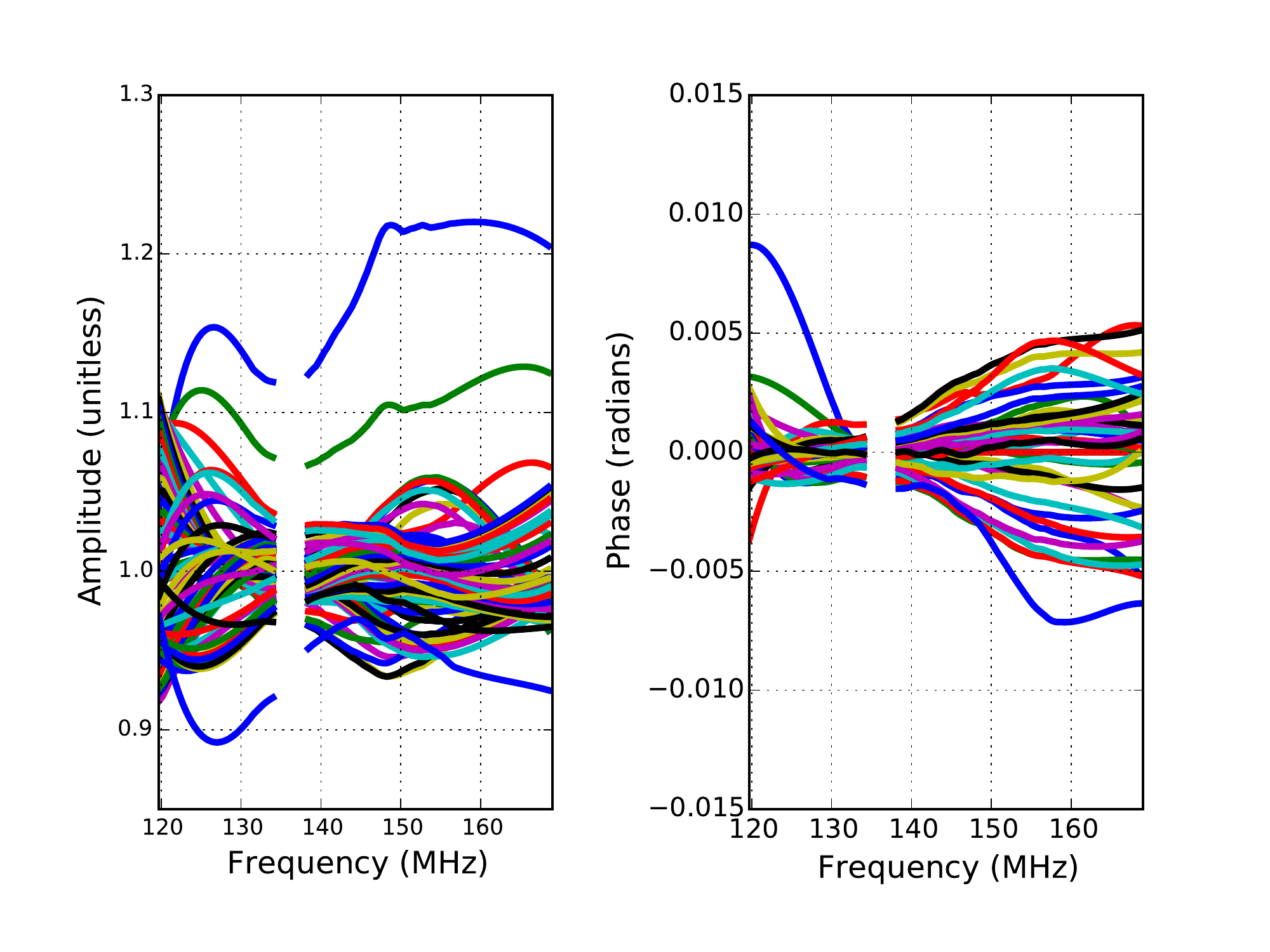}
	\caption{The bandpass gain solution derived from a snapshot of 10minutes observation (JD2456242.25733) of PAPER-64 elements. Amplitude (left) and phase (right) of the gain solutions against frequency. Each line is a different antenna. Discontinuity in the gain solution shows the flagged frequency channel that affected by RFI contamination.}
	\label{fig:antgain}
\end{figure}

However, the natural formalism of \corrcal{} would allow one to fit for gain solutions as a function of all frequency channels at once. It means that, instead of solving an independent set of equations for every single frequency, a visibility data $\d$ from different frequency channels put together in a longer vector as, 
\begin{eqnarray}
\label{eq:vec_vis2}
\d = \begin{bmatrix}
\operatorname{\mathbb{R}e}\{\mathrm{v({f_{1})}}\},\ \operatorname{\mathbb{I}m}\{\mathrm{v(f_{1})}\},\ \operatorname{\mathbb{R}e}\{\mathrm{v(f_{2}})\},\ \operatorname{\mathbb{I}m}\{\mathrm{v(f_{2}})\},\  \cdots 
\end{bmatrix}, \nonumber \\
\label{eq:d-matrix2}
\end{eqnarray}
where $\operatorname{\mathbb{R}e}\{\mathrm{v}(f_{1})\}$,  $\operatorname{\mathbb{I}m}\{\mathrm{v}(f_{1})\}$, $\operatorname{\mathbb{R}e}\{\mathrm{v(f_{2}})\}$, and $\operatorname{\mathbb{I}m}\{\mathrm{v(f_{2})}\}$ are the real, $\operatorname{\mathbb{R}e}$, and imaginary, $\operatorname{\mathbb{I}m}$,  parts of the  visibility measurements corresponding to frequency channel $f_{1}$ and $f_{2}$, respectively. Putting together visibility from all frequency channels in a longer vector like that enforce \corrcal{} to execute covariance information between frequencies. This imposes one to solve for smooth bandpass gain solutions for all frequencies as a single calibration step as depicted in Fig.~\ref{fig:antgain}. To some extent, the frequency structure of the gain solutions in Fig.~\ref{fig:antgain}  is comparable to the autocorrelation result in \citet{Li2019} and \citet{Byrne2019}.  Once these bandpass calibration solutions are found, we divide these solutions to the raw data for \corrcal{} calibration. 

The correlation calibration suffers from the same degeneracy problem as with redundant baseline calibration that presented in section \ref{sec:redundunt_cal}. Therefore, the gain solutions in \corrcal{} are degenerate, and these degeneracy issues will have to be dealt with. In this work, the overall amplitude degeneracy in \corrcal{} gain solutions is constrained by setting the average absolute value of the gains equal to one. Because we are applying \corrcal{} to data that has already been absolutely calibrated, we do not want to rescale the overall amplitude. To set the overall phase degenerate parameter $\Delta$, we define the reference antenna. We then set the overall phase by bringing the argument of the gain of the reference antenna equal to zero.  The phase gradient degeneracy issues are resolved using the known point sources information on the sky.  Then, we solve for phase gradient parameters by fitting them with the phase solutions of gain that have been obtained using sources information. That means,
 \begin{eqnarray}\label{eq:overall_phase}
 \chi^{2}_{\rm phase}=\sum_{p=1}^{N}\left({\rm Arg}
 [\hat{g}_{p}(f)]-\Delta_{x}b_{x,p}-\Delta_{y}b_{y, p}\right),
 \end{eqnarray}
 where $b_{x,p}$ and $b_{y, p}$ are the $x$ and $y$ components of the  baseline $b$ for antenna $p$.  Then, we subtract the fitted gradient parameters off from the phase calibration solutions across the array to set for phase gradient issues.

\subsection{Comparisons of \corrcal{} to other methods}
The \corrcal{} approach has a  distinct feature to alleviate issues like imperfections of array redundancy and miss-modeled sky catalog that \textit{redundant baseline} and \textit{sky-based} calibration techniques suffer from.  In \corrcal{},  sources with known positions can be included as a prior even if their fluxes are not precisely known, because the position of the sources is better known than their intensities during observation time and frequency.  This concept allows \corrcal{} to assume the source catalog is complete and well-modeled with precisely known source positions.  The method works well if the sky-modeling error could be mostly caused by uncertain/missed source fluxes in the \textit{sky-based} calibration approach. 

The covariance-based formulation of $\chi^{2}$ in \corrcal{} provides significant flexibility by relaxing the assumption of explicit redundancy to accounts for imperfections of array redundancy. For instance, if baselines in a quasi-redundant group are assumed to be alike but they are not perfectly alike, then the \corrcal{} method incorporates this offset from perfection in its formalism by suppressing the off-diagonal elements of covariance matrix within the group. However,  the redundant baseline method assumes that the baselines are perfectly redundant in the group,  and therefore all the off-diagonal elements of the covariance matrix in the group are exactly equal(e.g.~\refeq{eq:example_r}).

\section{The PAPER-64 Data Set}
\label{sec:data}
PAPER is a low-frequency radio interferometer experiment dedicated to probing the EoR through the measurements of the 21-cm power spectrum~\citep{Parsons2010}. It was the first 21-cm array experiment with a redundant array configuration, where antennas were arranged in a regular grid to generate multiple copies of the same baselines. The redundant baselines probe the same modes on the sky, resulting in significant improvement in the sensitivity of the 21-cm power spectrum,  thus, it is reasonable to use redundant calibration. The array was operating at the South African SKA site in the Karoo desert until its decommissioning in 2015. It was first deployed with 16 antennas and continued to grow, eventually increasing to 128 antennas in 2015. PAPER has 1024 frequency channels across the 100 MHz band. It observes between 100-200~MHz, corresponding to the 21 cm signal from redshifts 6-13, with a frequency resolution of $\sim$97 KHz. Data from the 32-elements PAPER array has been analyzed in~\citet{Parsons2014}, which results in one of the very first upper limits on the 21-cm power spectrum at $z=7.7$. Analyses of the data set from the 64-elements PAPER array (hereafter PAPER-64) in \citep{Cheng2018,  Kolopanis2019} have placed significant upper limits on the power spectrum amplitude of 21 cm signal.  

\begin{figure}
\centering
   \includegraphics[width=\columnwidth]{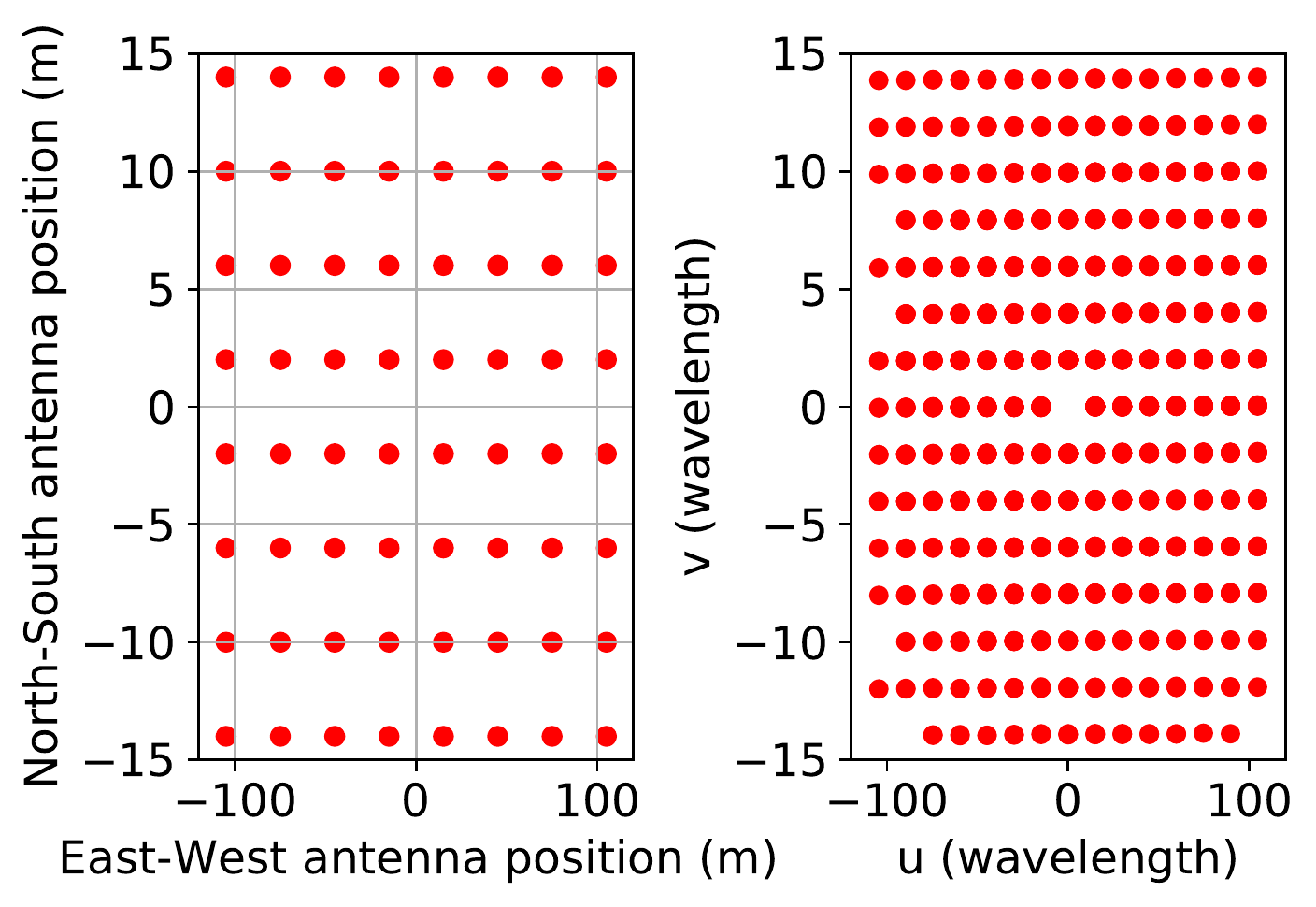}
   \caption{Antenna position (left) and UV coordinate  of the PAPER-64 array. The North-South and East-West antenna coordinates are in meter, and the UV coordinates are in wavelengths at 150\,MHz (right).}
   \label{fig:antpos}
\end{figure}

We will perform a recalibration of the PAPER-64 data set taken from~\citet{Ali2015} with \corrcal{}. This data set has been calibrated with redundant calibration using the \texttt{OmniCal} software, and then absolute calibrated. The absolute calibration has been made by fitting to the bright sources such as Fornax A, Pictor A, and the Crab Nebula to set the overall phase. Using the Pictor A calibrator, the overall amplitude in this data has been fixed. The detailed discussions about the absolute calibration stage for this data are presented in~\citet{Ali2015}.  The data set has dual polarization products.  We use one night of observations taken between 10 November 2012 (JD2456242.17382) and 11 November 2012 (JD 2456242.65402).  As depicted in Fig.~\ref{fig:antpos}, the PAPER-64 array system comprises 64 dipole antennas arranged on a regular grid spacing with the total number of baselines equal to $64\times(64-1)/2=2016$ and thus measure 2016 visibilities. For our calibration, we use all baselines except those associated with three antennas which were flagged during the observation due to known spectral instability. The observation parameters of this data set are summarized in Table~\ref{Tab:paper_summary}.  We direct readers to~\citet{Ali2015} for more details of observation strategies, flagging of radio frequency interference (RFI), channelization of data, antenna cross-talk minimization, and other technical information of the PAPER array. Our method proposes a new possibility to fully gauge the systematics and noise of radio interferometry.

\begin{table}
\caption{Summary of observational parameters and their values used in this work}
\centering
\begin{tabular}{|p{3.5cm}|p{3.5cm}|}
\hline
Parameter& Value\\
\hline
PAPER array location & $\mathrm{30.7^{o} S, 21.4^{o}\,E}$\\
Observation dates & 10 Nov 2012-11 Nov 2012 \\
Observing mode& drift-scan\\
Time resolution&$\sim$42.9 seconds\\
Frequency range& 120\,MHz-168\,MHz\\
Frequency resolution& 0.49\,MHz\\
Number of antenna& 61\\
Field-of-view&60$^{\circ}$\\
Visibility polarization & $XX$,$YY$ \\
Shortest baseline & 4m\\
Longest baseline & 212m \\
\hline
\end{tabular}
\label{Tab:paper_summary}
\end{table}
\section{Results}
\label{sec:results}
\subsection{Post-\corrcal{} spectral structure of data}
\begin{figure}
	\centering
	\includegraphics[width=0.4\textwidth]{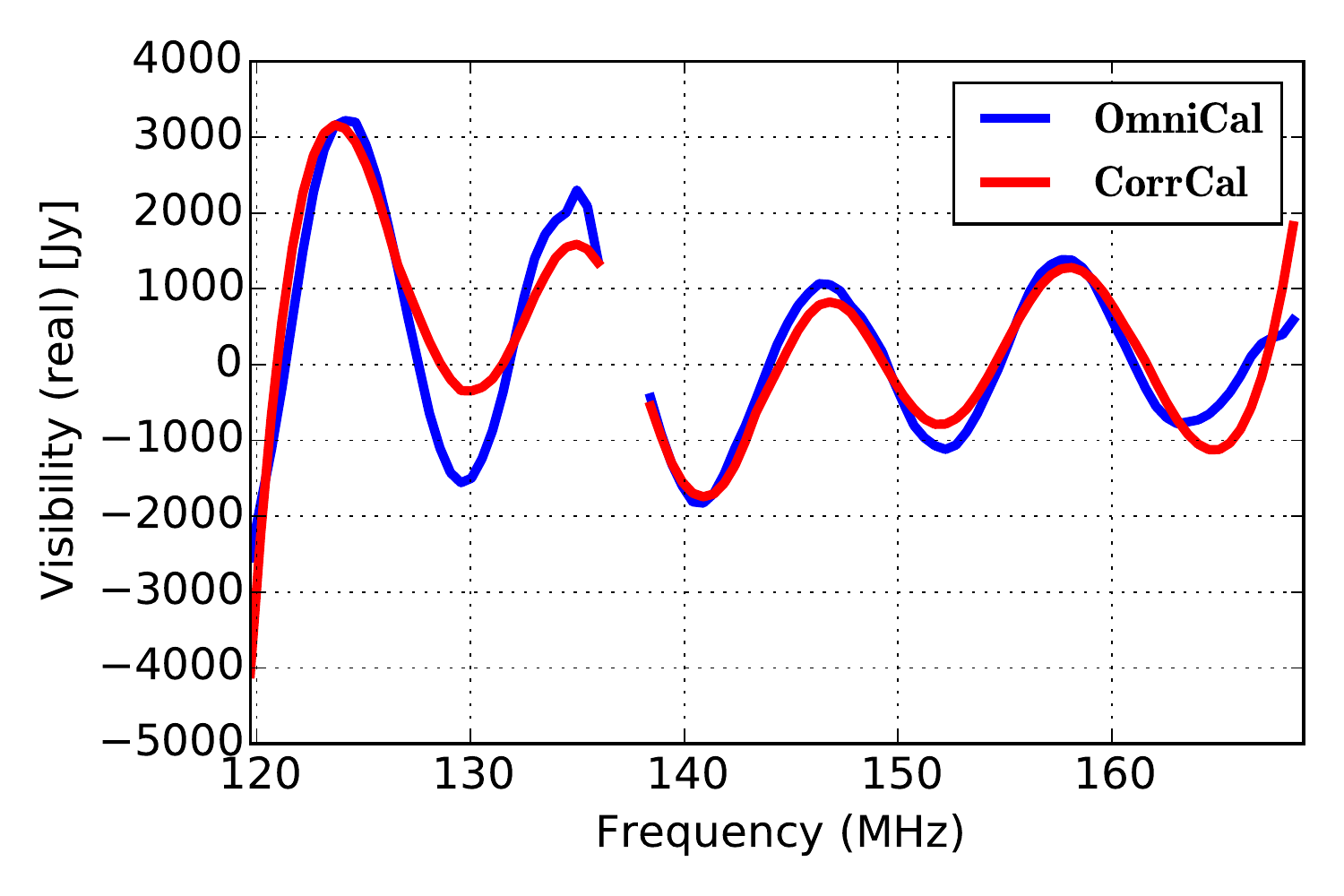}
	\caption{The real part of visibility data (JD2456242.25733) as a function of frequency at a single timestamp for an EW baseline of length about $30$\,m for both calibrations approaches. Line discontinuity corresponds to the flagged frequency channels. The spectral structure looks slightly reduced after \corrcal{} (red curve).}
	\label{fig:visplot}
\end{figure}

After \corrcal{} implementation, we have tried to compare the structure of data as a function of frequency. Figure~\ref{fig:visplot} depicts the real part of visibility against frequency for a single integration time of a given baseline. The careful examination of the plot shows that the real-visibility gets spectrally smoother (red curve)  after \corrcal{} re-calibration. It signifies the reduction of the spectral structure of the visibility measurements after re-calibration. Considering that, in the following section, we present the effect of re-calibration on the delay transformed power spectra ~\citep{Parsons2009}.
\subsection{Delay-space power spectra}
The statistical analysis of the Fourier mode of the power spectrum ($k_{\perp},k_{\parallel}$) for low-frequency observations is a powerful tool to probe the EoR~\citep{Morales2004}. This method  shows the existence of  distinctive regions that dominated by 21-cm signal and foreground contaminants in a 2D line-of-sight ($k_{\parallel}$) and transverse comoving $\left( k_{\perp}=\sqrt{k^{2}_{x}+k^{2}_{y}}\right)$ Fourier plane. The spectrally smooth foreground is expected to contaminate the region of low $k_{\parallel}$ values, while the vast majority of $k_{\parallel}$ values are free from these contaminants~\citep{Liu2014a}. Nevertheless, several studies have shown that foreground contamination may go further to the higher $k_{\parallel}$ values because of chromatic interaction of interferometry  with intrinsically smooth foreground \citep{Parsons2012a, Morales2012, Dillon2014,Vedantham2012,Thyagarajan2013, Liu2014a,Liu2014b}. This effect leads to the distinctive wedge-like structure in $k_{\perp}$ and $k_{\parallel}$ Fourier space, leaving the foreground-free EoR window beyond the wedge (see results in Fig.~\ref{fig:wedge} and discussions there).

\begin{figure*}
 \centering
 \includegraphics[width=0.7\textwidth]{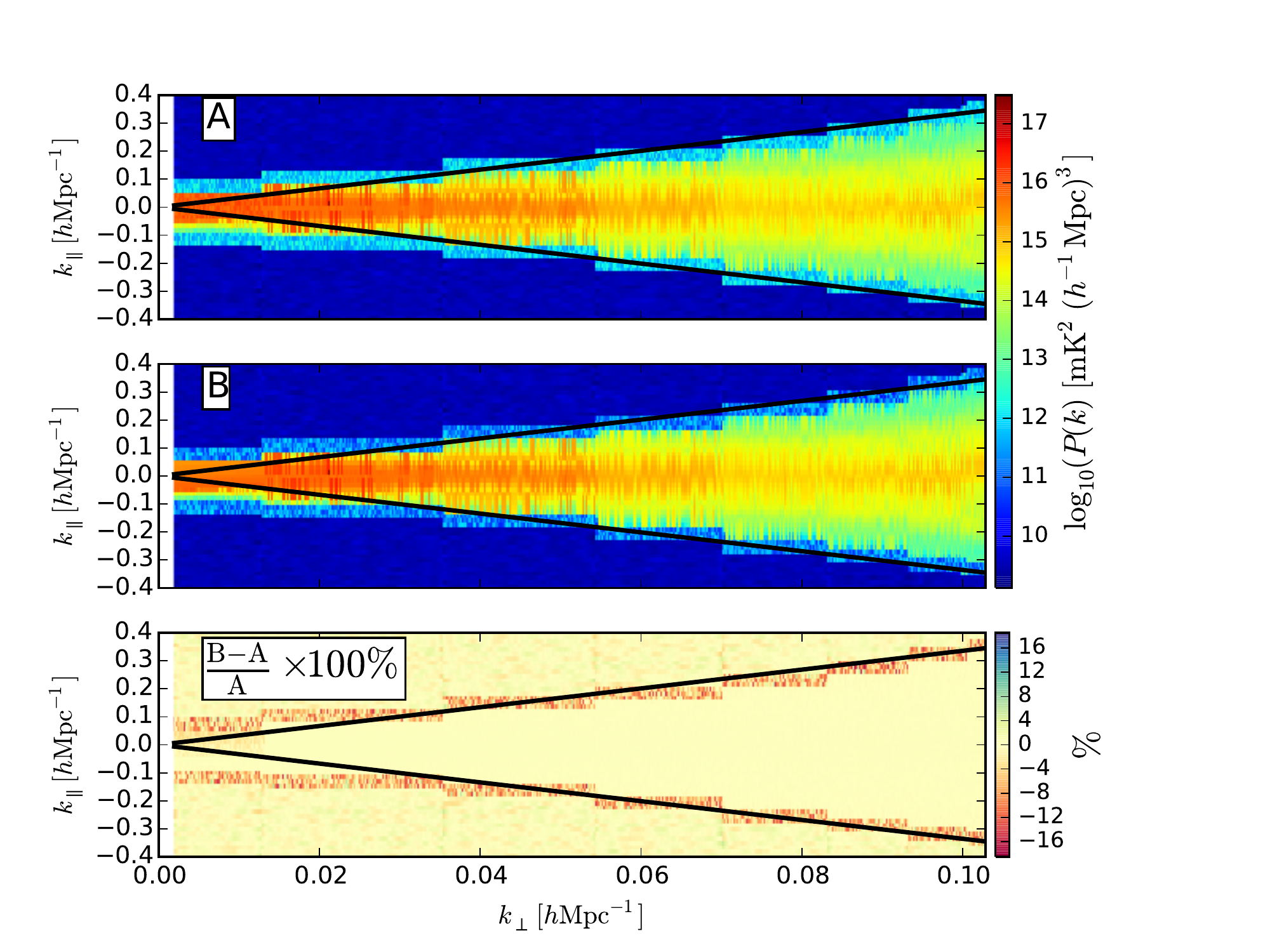}
 \caption{Delay power spectra on either side of $k_{\parallel}$ direction. Foreground wedges are confined under the region bounded by the black line or the horizon line. (A) the delay power spectra for data already calibrated with \texttt{OmniCal}, (B) power spectra after \corrcal{} recalibration,  and the lower panel displays linear scale percentage deviation. The negative percentage deviation in power at the bin right on the horizon shows that the wedge has become tighter after \corrcal{} recalibration.}
 \label{fig:pitchfork}
\end{figure*}

To convert the measured visibility to 2D power spectrum in $k_{\parallel}$ and $k_{\perp}$ space, the per-baseline delay transform method is used. The technique was first introduced by~\citet{Parsons2009}. As thoroughly presented in~\citet{Parsons2012a}, the delay transformed visibility is obtained by inverse Fourier transforming of each baseline along the frequency axis. The method allows the smooth foreground to localize at the central region between the negative and positive geometric delay limits, and any spectrally unsmooth structure such as 21-cm signal may spread across delay axis~\citep{Parsons2012a}. We will review the mathematical formalism of the per baseline delay-transform method in Appendix~\ref{sec:delay}. In our analysis, we use Stokes $I$ visibilities\footnote{Note that these are not true Stokes $I$ would rather the average of $XX$ and $YY$ visibilities.(see Sec.~\ref{sec:delay})} taken by PAPER-64 array from Julian date 2456242.17382 to 2456242.65402 (2012 November 10-11) to form the 2D power spectrum using delay transform method, a total of about 9 hours observation over frequencies 140 to 160MHz.

Following~\citet{Parsons2012a} and~\citet{Pober2013}, we apply the Blackman-Harris window function during the delay-transform stage to minimize the foreground leakage caused by the finite bandwidth of frequency.  This technique allows the EoR window to be mostly dominated by 21-cm brightness temperature by reducing the foreground spectral leakage to the smallest possible amount~\citep{Parsons2012a, Pober2013}. We employ the 1D CLEAN algorithm on the delay transformed data to reduce the effect introduced by RFI flagging and to increases the resolution in delay space~\citep{Pober2013}. We then cross-multiply the consecutive integration time to avoid the noise bias. After these stages, we form the cylindrical delay power spectrum in $k$ space using~\refeq{eq:28}.

Figure~\ref{fig:pitchfork} shows the delay power spectra where the positive and negative values of $k_{\parallel}$ have not been averaged together. This form of power spectra is expected because emissions from the negative and positive delay directions are proportional to the negative and positive $k_{\parallel}$ respectively. The delay power spectra from emissions near-zero delays (where $k_{\parallel}\sim 0$)  are confined within the central region of the 2D  delay power spectra. It corresponds to foreground emissions from the peak primary beam response of the instrument.

 However, from Fig.~\ref{fig:pitchfork}, we see that the strength of power spectra around zero delay decreases as one moves from the lower $k_{\perp}$ to higher $k_{\perp}$ regions. 
 Furthermore, for both calibration approaches, more emission goes beyond the horizon limit for shortest baselines (smallest $k_{\perp}$ values). This effect is because shorter baselines resolve out less of the Galactic synchrotron so that emission will be brighter and sidelobes extend further in~$k_{\parallel}$~\citep{Pober2013}.
 
The black diagonal line represents the horizon limit determined by baseline length and the source location relative to the main field of view of the array, marking the boundaries of the foreground wedge. Its analytic expression has been defined in~\refeq{eq:30}.  The spectral-smooth emissions are supposed to be confined within that limit, and any emissions which are intrinsically unsmooth with frequency are moving beyond the horizon limit. The calibration error plays a non-negligible role in this regard in imparting the spectral structure to emissions that were originally spectrally smooth as discussed earlier. This unnatural spectral structure, imparted due to calibration error or other effects,  will affect the EoR window by scattering power beyond the wedge limits. To see \corrcal's impact on the delay power spectra, we compare foreground power in a wedge region of the power spectrum, and outside of it in which the 21-cm signal is dominating by taking the percentage deviation between power after \corrcal{} and \omnical{}, shown in the bottom panel of  Fig.~\ref{fig:pitchfork} . It is apparent that the power spectra inside and outside the wedge region are comparable for both calibration approaches.  However, we observe a consistent drop in power at the bin right on the horizon after \corrcal{}.  Most importantly, this is consistent with some of the analyses done with MWA in  \citep{Li_2018,Li2019,Zhang2020} that demonstrate redundant calibration's ability to make small improvements in the modes nearest (but not inside) the wedge. This could be more suggestive of a reduction in the spectral structure of the data after re-calibration with \corrcal{}.
\begin{figure*}
	\centering
	\includegraphics[width=0.8\textwidth]{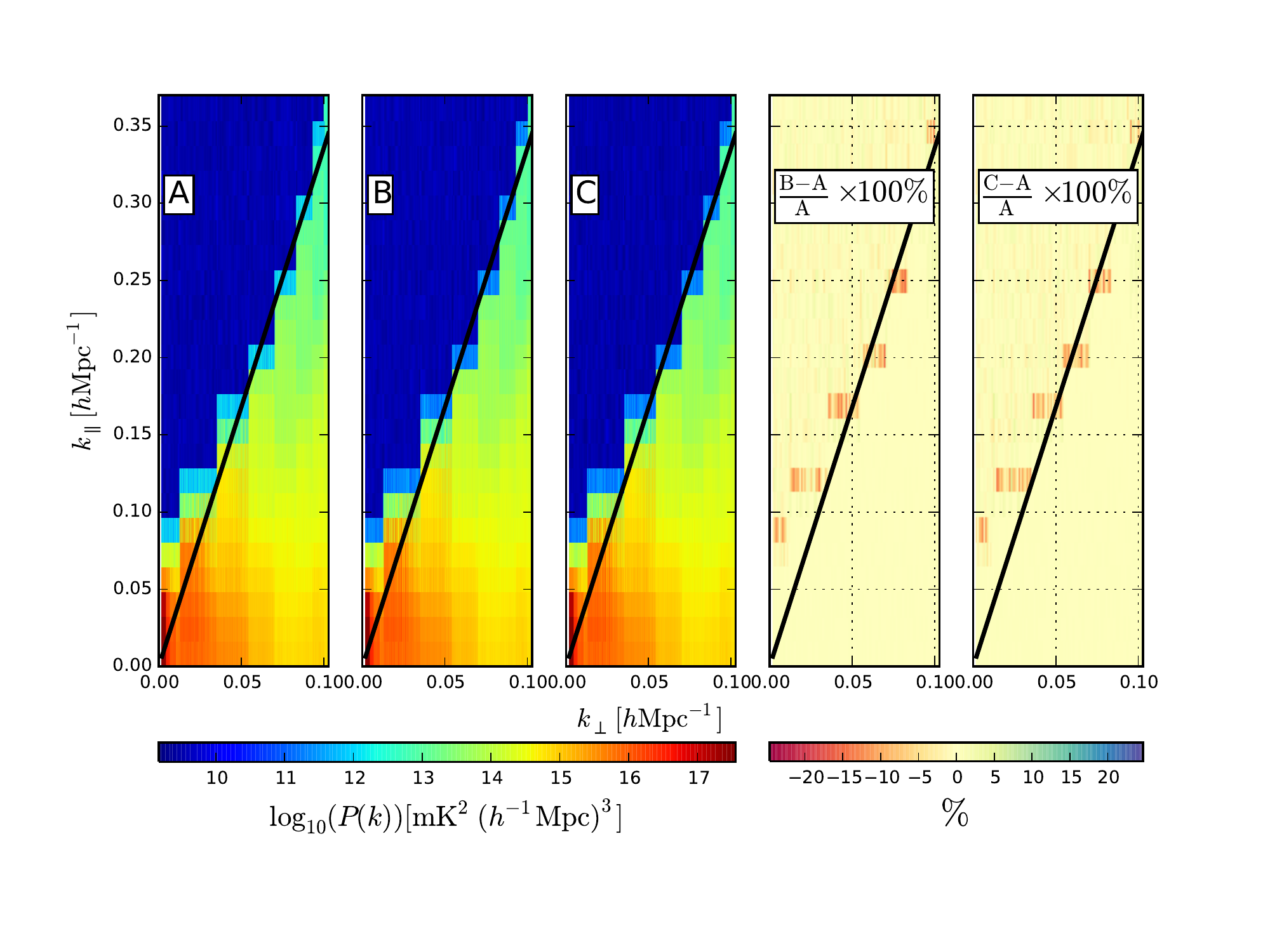}
	\caption{Same as Fig.~\ref{fig:pitchfork} but averaged and folded over the delay centre to obtain the ``wedge''-shaped delay power spectra. From left to right: (A) The  power spectra generated from data already calibrated with \texttt{OmniCal} package; (B) the power spectra obtained after data have recalibrated using without source information in \corrcal{}; (C) power spectra after recalibration using source information in \corrcal{}. The right two panels show the  linear scale percentage deviation without and with sources information in \corrcal{} formalism respectively.  The black diagonal line marks the horizon limit ($\tau_{\rm H}$).}
	\label{fig:wedge}
\end{figure*}

The well-known wedge-shaped foreground power spectra from our analysis are shown in Fig.~\ref{fig:wedge}. We generate this form of power spectra by averaging power from both positive and negative  $k_{\parallel}$ directions and folding it over the centre of the zero-delay line (where $k_{\parallel}=0$) of Fig.~\ref{fig:pitchfork}. A similar feature of power spectra has been observed from PAPER-64 data in~\citet{Pober2013}, as displayed in Figure 3 of their work. Moreover,~\citet{Kohn2016} have reported a wedge-like structure from the analysis of PAPER-32 data. In our analysis,  Fig.~\ref{fig:wedge}(A) and (B), respectively, displaying the wedge-shaped power spectra to data already calibrated using \texttt{OmniCal} and to data after \corrcal{} recalibration. The middle panel in  Fig.~\ref{fig:wedge} (C) shows power spectra after data have been calibrated using point sources information in \corrcal{}. For each case, the wedge extends to higher values of $k_{\parallel}$ at a longer baseline ($k_{\perp}$) regime, demonstrating how the theoretical wedge limit increases as a function of baseline length. This effect is also due to sources located far away from the pointing centre of the primary beam of the instrument appears at higher delays, which corresponds to higher $k_{\parallel}$.  The relative deviation of power in percent between \corrcal{}  and \omnical{}  calibration approaches are shown in the third (\corrcal{} without sky sources information ) and fourth  (\corrcal{} with sky sources information) panels of Fig.~\ref{fig:wedge} from left to right , respectively. The negative percentage deviation at the bin right on the horizon in  $\mathrm{[(B-A)/A]\times 100\%}$ and $\mathrm{[(C-A)/A]\times 100\%}$ of Fig.~\ref{fig:wedge} indicating that the power spectra around the wedge limit have become slightly tighter after \corrcal{} recalibration. This effect is not considerably saturated if we include the sources' information in  \corrcal{}  formalism as shown in the right panel of the figure.

\begin{figure*}
  \centering
  \includegraphics[width=18cm]{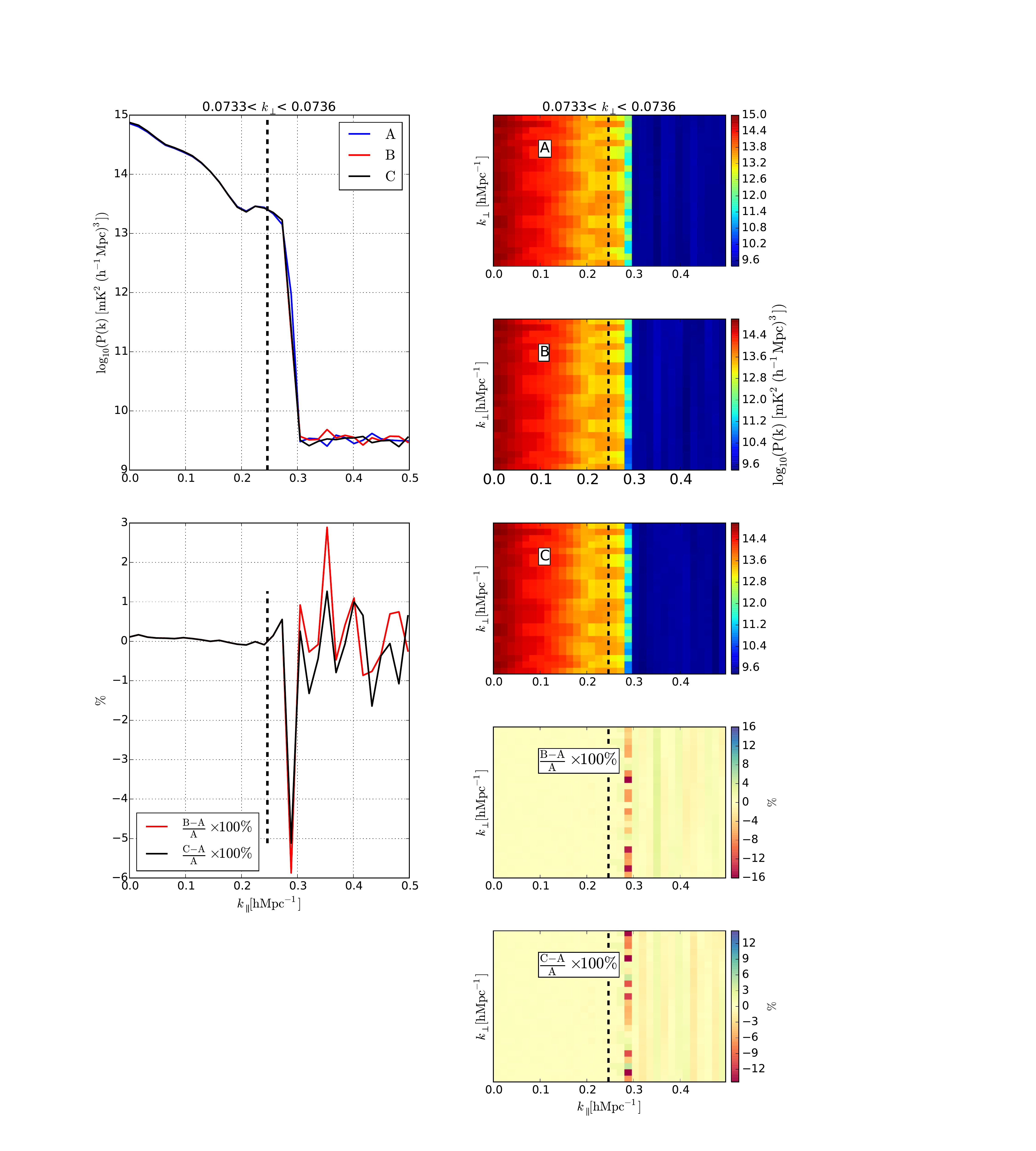}
  \caption{{{\it Right}--Two-dimensional power spectra after (A) \texttt{OmniCal}, (B) \corrcal{} recalibration without sources info, (C) \corrcal{} recalibration with sources information, the lower panels are the percentage deviation of  B to A and C to A,  respectively. The black dashed vertical line is representing the horizon ($\tau_{\rm H}$) limit. \it Left}--The averaged slice of power spectra  from $k_{\perp}=0.0733~h\,{\rm Mpc}^{-1}$ to $k_{\perp}=0.0736~h\,{\rm Mpc}^{-1}$ to both calibration schemes. The top panel is the one-dimensional average power spectrum as a function of $k_{\parallel}$  for a portion of baselines ($\sim150\rm m$) to both calibration strategies. The bottom panel is the percentage deviation.}
  \label{fig:slice}
\end{figure*} 

The R.H.S panel of Fig.~\ref{fig:slice} depicts a portion of 2D power spectra for a given group of baselines whose lengths are about $150$\,m. We average over $k_{\perp}$ axis over the range of $0.0733\,h\,{\rm Mpc}^{-1} < k_{\perp} < 0.0736\,h\,{\rm Mpc}^{-1}$ to generate the corresponding 1D power spectra that shown on the L.H.S of Fig.~\ref{fig:slice}. From this figure,  we observe that there is a steep fall-off of the power spectrum right away the horizon limit as expected, showing that the foreground power confined within the slice of the wedge is brighter than the 21-cm signal by 4-5 orders of magnitude. This sharp fall-off follows the same trend for both calibration schemes. However, the inclusion of sky sources statistics as a prior during \corrcal{}  recalibration step doesn't show the significant change in the power spectra,  as shown on the L.H.S of Fig.~\ref{fig:slice}.  

\section{Conclusions}
\label{sec:conclusions}

In this work, we used a new calibration scheme (\corrcal)  to recalibrate the PAPER64 observations. Our new calibration scheme boils down to the covariance-based calculation of $\chi^{2}$, allowing one to include the relaxed array redundancy and sky information into its formalism.  This formalism also provides space for cross-frequency bandpass calibration. Furthermore, the inclusion of known sources information in \corrcal{}  breaks the statistical isotropy of the sky,  which sets the overall phase gradient degenerate parameter across the array. We have compared the spectral smoothness of the observed data before and after \corrcal's application. We observed that the spectral structure of the visibility data after  \corrcal{} becomes slightly reduced, for instance, as shown in ~Fig. \ref{fig:visplot}.  The significant factor for this improvement could be the  \corrcal{} natural formalism to carry out the cross-frequency bandpass calibration.

We implemented the delay-transform method~\citep{Parsons2009, Parsons2012a} to filter out smooth foregrounds from the spectrally structured weak 21-cm signal in 2D Fourier $k-$space.  The comparison results in the delay space show that the power spectra around the foreground wedge limit have been reduced by about $6\%$ after the \corrcal{} re-calibration. Nevertheless, this effect does not significantly improve the power spectrum in any bin around the horizon limit of the foreground wedge and further investigation will be needed to test the algorithm effectiveness and its implementation methods. The current numerical tests lay the foundation of calibration radio instruments using correlated signals on the sky. Our future work will include possible beam shape errors, antenna pointing errors, and position errors, and improve the \corrcal{} method by using sky and antenna simulation (e.g., {\tt pyuvsim}) and compare it with sky-based calibration and redundant baseline calibration. In conclusion, the \corrcal{} may be an alternative calibration method for radio interferometry, which potentially helps to probe the EoR using the 21cm signal. 

\section*{Acknowledgements}
We would like to acknowledge the PAPER collaboration for providing us data that have been used in this work. Y.Z.M. acknowledges the support of NRF-120385, NRF-120378, and NSFC-11828301. P.K. acknowledges funding support from the South African--China Collaboration program in Astronomy at the University of KwaZulu-Natal.  

\section*{Data Availability}
The data underlying this article will be shared on reasonable request to the corresponding author.


\bibliographystyle{mnras}
\bibliography{bibliography} 



\appendix

\section{Delay Spectrum}
\label{sec:delay}
After data with $XX$ and $YY$ linear polarisation being recalibrated using \corrcal{} separately,  we form the Stokes $I$ parameter by combining each linear polarisation following~\citet{Moore2013}:
 \begin{equation*}
 \label{eq:stokes}
     V_{\mathrm{I}}=\frac{1}{2}(V_{XX}+V_{YY}),
 \end{equation*}
where $V_{XX}$ and $V_{YY}$, respectively, are visibilities from $XX$ and $YY$ linear polarisation. We then use $V_{\mathrm{I}}$ (Stokes $I$ polarisation visibility)  to form the power spectra using a delay-transform approach following that of~\citet{Ali2015}.

Approximating the interferometric visibility equation that is defined in~\refeq{eq:visibility_eq} over the  flat-sky model, we have
\begin{eqnarray}\label{eq:24}
\d(u,v,f)=\int \textrm{d}l\, \textrm{d}m\, I(l,m,f)A(l,m,f)e^{-2\pi i(ul+vm)},
\end{eqnarray}
where $(l,m)$ are direction cosines  of a unit vector $\ur$, which pointing to the source on the sky. These sky-plane direction  cosines  have the corresponding antenna-plane Fourier dual $u$ and $v$, respectively,  that is defined in Section~\ref{sec:cov_matrices}. Following the representation of~\citet{Parsons2009}, we can rewrite~\refeq{eq:24} in terms of geometric time delay $\tau_{\mathrm{g}}$  for a single baseline $\b$ as
\begin{eqnarray}\label{eq:25}
    \d_{\b}(f)=\int \textrm{d}l\, \textrm{d}m\, I(l,m,f)A(l,m,f)e^{-2i\pi f\tau_{\rm g}},
\end{eqnarray}
where 
\begin{eqnarray}\label{eq:26}
\tau_{\mathrm{g}}=\frac{1}{c}\b\cdot\ur=\frac{1}{c}(b_{x}l+b_{y}m),
\end{eqnarray}
 with $\b=(b_{x},b_{y})$, $(u,v)=f\mathbf{b}/c$ and $\ur=(l,m)$. Here, $b_{x}$ and $b_{y}$ are projected baseline length in meter along east and north direction in the antenna plane, respectively. Note that $\tau_{\rm g}$ in~\refeq{eq:25} shows the time-shifting of signal arriving one antenna relative to other due to source location on the sky with respect to the baseline vector orientation in the antenna plane.

Delay transform technique begins by converting the frequency spectrum of the visibility into the delay spectrum through application of Fourier transform on the visibility along its frequency direction. Applying the windowing function $W(f)$ over~\refeq{eq:25} and  then Fourier transforming it  along its  frequency axis, the delay transformed visibility takes a form
 \begin{eqnarray}
 \label{eq:27}
  \tilde V_{\mathrm{b}}(\tau) &=& \int_{B} \textrm{d}f\, W(f)v^{\mathrm{mea}}_{\rm b}(f)e^{2\pi if\tau}\nonumber \\
    &=& \int_{B} \textrm{d}f\, W(f)\left[\int\textrm{d}l\, \textrm{d}m\, I(l,m,f)A(l,m,f)e^{-2\pi if\tau_{\rm g}}\right] \nonumber \\
    & \times & e^{2\pi if\tau} 
    \nonumber \\
    &=& \int_{B} \textrm{d}l\,\textrm{d}m\, \tilde{W}(\tau)*\tilde{I}(l,m,\tau)*\tilde{A}(l,m,\tau)*\delta_{\rm D}(\tau-\tau_{\rm g}), \nonumber \\
 \end{eqnarray}
 where the window function $W(f)$ chosen by the data analyst is used to control the quality of delay spectrum as suggested in~\citet{Vedantham2012} and \citet{Thyagarajan2013}, $B$ is the bandwidth of observation over which  the integration has to be carried out, the delay parameter $\tau$ is the Fourier dual of $f$ in unit of time, $\tilde{V_{\mathrm{b}}}(\tau)$ is the delay transformed visibility observed by a baseline $b$, * stands for convolution. Hence, from~\refeq{eq:27} one can clearly sees that there is mapping between  point sources on the sky in $\tilde{V}_{\mathrm{b}}$ space and the Dirac-delta function $\delta_{D}$, convolved by Fourier transform of window function $\tilde{W}(\tau)$, sky flux density $\tilde{I}(l,m,\tau)$, and the antenna beam response $\tilde{A}(l,m,\tau)$, as shown in \citet{Parsons2012a}. Analytically, the delay formalism that relates the spatial power-spectrum $P_{21}(k_{\parallel},k_{\perp})$ of 21-cm signal from the EoR fluctuations to the delay-transformed visibility $\tilde{V}$ has been  derived in~\citet{Parsons2012a} as
 \begin{eqnarray}\label{eq:28}
     P_{21}(k_{\parallel},k_{\perp})\approx\tilde{V}_{21}^{2}\left(\frac{\lambda^{2}}{2k_{\mathrm{B}}}\right)^{2}\frac{X^{2}Y}{\Omega B},
 \end{eqnarray}
 where $\lambda$ is the observing wavelength, $k_{\mathrm{B}}$ is the Boltzmann constant, $B$ is the observing bandwidth, $\Omega$ is the solid angle of the primary-beam of the antenna, $X$ and $Y$ are cosmological scalars  which convert observed angles and frequencies into $h\,\mathrm{Mpc^{-1}}$ as calculated in~\citet{Parsons2012a}.

Following the method in~\citet{Pober2013}, the successive time stamp in delay-transformed visibilities has been cross-multiplied to avoid the noise-bias. Using the formalism in~\citet{Kohn2016}, it can be written as 
 \begin{eqnarray}\label{eq:29}
  \tilde{V}_{21}^{2}\approx|\tilde{V}_{21}(\tau,t)\times\tilde{V}_{21}(\tau,t+\Delta t)e^{i\theta_{\rm zen}(\Delta t)}|^{2},
 \end{eqnarray}
here $\Delta t\approx$ 42.9 seconds (a time resolution of PAPER-64 data) and $\theta_{\rm zen}(\Delta t)$ is the zenith re-phasing factor. 
 The line-of-sight $k_{\parallel}$,  and transverse co-moving $k_{\perp}$ components of $k$ in terms of the instrumental and cosmological parameters have shown in~\citet{Morales2004},
\begin{equation*}
   k_{\perp}=\frac{2\pi |\mathbf{u}|}{D_{\mathrm{m}}(z)},  
\end{equation*}

\begin{equation*}
  k_{\parallel}\approx \eta\frac{2\pi H_{0}f_{10}E(z)}{c(1+z)^{2}},
\end{equation*}
where $|\mathbf{u}|=\sqrt{u^{2}+v^{2}}$, $\eta$ is the Fourier conjugate of $f$ which used to denote the spatial frequency along the line-of-sight and it has the unit of time, $f_{10}=1420$\,MHz is the rest frequency of 21-cm emission from cosmic reionization, $c$ is the speed of light in free space,  $D_{\mathrm{m}}(z)$ is the transverse comoving distance at the reference redshift $z$, the present day Hubble constant $H_{0}=70\,{\rm km}\,{\rm s}^{-1}\,{\rm Mpc}^{-1}$, the function $E(z)=\sqrt{\Omega_{\mathrm{m}}(1+z)^{3}+\Omega_{k}(1+z)^{2}+\Omega_{\Lambda}}$. Here the cosmological density parameters $\Omega_{\mathrm{m}}$, $\Omega_{\Lambda}$ and $\Omega_{k}$ are total matter density, the dark energy density and curvature, respectively. The geometric delay $\tau_{\mathrm{g}}$ in~\refeq{eq:26} sets the maximum delay limit beyond which no emissions enter the interferometry involving that baseline. This limit will be attained when the angle between $\b$ and $\ur$ is set to be zero. Such an alignment to both vectors is  achieved, the maximum delay is commonly referred to "horizon limit ($\tau_{\rm H}$)". Theoretically, over the  $(k_{\perp},k_{\parallel})$ cylindrical space, it can be determined from
\begin{eqnarray}\label{eq:30}
     k_{\parallel}=\frac{H_{0}E(z)D_{\rm m}(z)}{c(1+z)}k_{\perp}.
\end{eqnarray}
As an exhaustively discussed in~\citet{Parsons2012a}, any strictly smooth foreground emissions with power-law spectra are confined within the narrow range of horizon limits such that $-\tau_{\rm H}\leq\tau\leq\tau_{\rm H}$. Non-smooth power either intrinsic to the sky or instrumental effects scattered beyond the horizon limit as sidelobes of the convolving kernel of sky and beam,  $\tilde{I}*\tilde{A}$, which broadens the footprint of  $k_{\parallel}$ modes in cosmological $k$ space.  This kernel is narrow for smooth spectrum foreground sources and confined within the wedge.  However, flux beyond the horizon limit is representing the power-spectrum of unsmooth emissions such as 21-cm signal. Therefore, regions in a wide range of $k_{\parallel}$ modes are considerably affected by the 21-cm signal if one keeps the frequency response of the instrument smoother. More of these modes are free from foreground contamination on shortest  baseline, i.e., at smallest $k_{\perp}$ regions \citep{Datta2010, Vedantham2012, Parsons2012a, Pober2013}.
\section{Determining the EoR signal loss in \corrcal{} analysis} 
To determine whether our analysis methods are lossless or not in the EoR power spectrum estimate, first, we simulate visibility data using HERA simulation pipeline (\texttt{hera\_sim}\footnote{https://github.com/HERA-Team/hera\_sim.git}). For this simulation,  we use the analytic airy beam model (the same for all antenna elements), the simulated antenna gains, the  GaLactic and Extragalactic All-sky MWA (GLEAM) sky sources\citep{Hurley2017}, the simulated EoR signal using \texttt{hera\_sim}. Then,  to quantify our new calibration approach does not lead to the EoR power spectrum signal suppression, we propagate simulated data into the \corrcal{} for calibration. We then compute the 2D power spectrum for both simulated and recovered data after \corrcal{} calibration using the delay spectrum approach described in Appendix \ref{sec:delay}.  ~\reffg{fig:recovered}  shows the delay space power spectra for both simulated and recovered data. From the figure,  it is apparent that the recovered power spectra after \corrcal{} are comparable to the expected one, confirming that \corrcal{} calibration method does not lead to significant signal loss.
\begin{figure*}
	\centering
	\includegraphics[width=15cm]{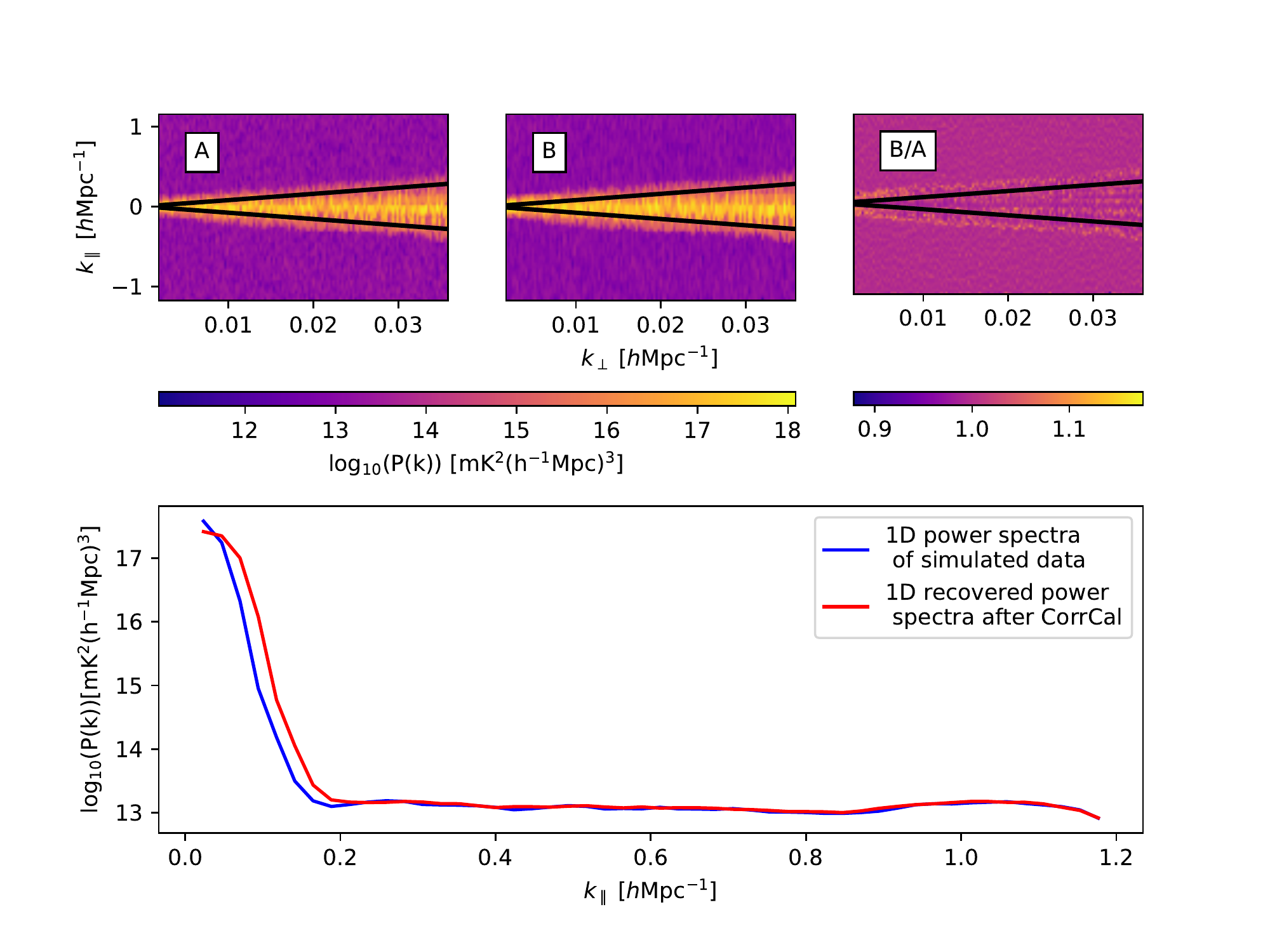}
	\caption{ {\it Top}--(\textbf{A}) Expected delay power spectra, (\textbf{B}) recovered power spectra after \corrcal{} calibration , (\textbf{C}) the power spectra ratio of  \textbf{B} to \textbf{A}. The solid black dashed line is the horizon limit, $\tau_{\rm H}$. {\it Bottom}--Averaged 1D power spectra  as a function of $k_{\parallel}$  for a portion of baselines in the $\sim 30~\rm m$ group. }
	\label{fig:recovered}
\end{figure*} 


\bsp	
\label{lastpage}
\end{document}